\documentclass[a4paper,UKenglish,cleveref,autoref,thm-restate]{lipics-v2021}

\hideLIPIcs  
\usepackage{float}
\usepackage{placeins}

\usepackage{tikz}
\usetikzlibrary{
  positioning, shapes.geometric, arrows.meta, 
  decorations.pathreplacing, backgrounds, fit
}

\usepackage{booktabs}
\usepackage{pifont} 
\usepackage{longtable}
\usepackage{amsmath,amssymb}
\usepackage{enumitem}
\usepackage{multicol}
\usepackage{longtable}
\usepackage{xcolor}

\usepackage[strings]{underscore}

\usepackage{tikz}
\usetikzlibrary{arrows.meta}
\tikzset{>={Straight Barb[angle'=80, scale=1.1]}}

\usepackage{fontawesome5}
\usetikzlibrary{positioning}

\usepackage{pifont}

\usepackage{pdflscape}

\usepackage{tikz}
\usepackage{tikz}
\usetikzlibrary{positioning, shapes.geometric, arrows.meta}

\newcommand{\cmark}{\ding{51}} 
\newcommand{\xmark}{\ding{55}}

\usepackage{algorithm}
\usepackage[noend]{algpseudocode}

\newcommand{\Thetabestfixedold}{\Theta_{\text{best\_fixed\_old}}}
\newcommand{\Thetabestfixedcurrent}{\Theta_{\text{best\_fixed\_current}}}

\setlist{nosep}

\usepackage{amsmath,amssymb}
\usepackage{longtable} 
\usepackage{array}

\newcommand{\Moffchain}{M_{\text{off-chain}}}

\newcommand{\Thetafixed}{\Theta_{\text{fixed}}}

\newcommand{\Pnewonchainfixed}{P^{(\text{new, on-chain})}_{\text{fixed}}} 
\newcommand{\Thetabestfixed}{\Theta_{\text{best\_fixed}}}

\newcommand{\Thetainferfixed}{\Theta_{\text{infer\_fixed}}}
\newcommand{\CPoIm}{C_{\text{PoIm}}}
\newcommand{\CInfer}{C_{\text{Infer}}}

\newcommand{\Dtest}{\mathcal{D}_{\text{test}}} 

\newcommand{\Sw}{S_w}
\newcommand{\Sm}{S_m}
\newcommand{\Xsample}{X_{\text{sample}}}
\newcommand{\Xsamplefixed}{X_{\text{sample\_fixed}}}
\newcommand{\Yoffchain}{Y_{\text{off-chain}}}
\newcommand{\Yonchain}{Y_{\text{on-chain}}}

\newcommand{\tofixed}[2]{\text{to\_fixed}(#1, #2)}
\newcommand{\fromfixed}[2]{\text{from\_fixed}(#1, #2)}
\newcommand{\integercast}[1]{\text{integer\_cast}(#1)}

 \title{On-Chain Decentralized Learning and Cost-Effective Inference for DeFi Attack Mitigation}

\author{Abdulrahman Alhaidari}{School of Computing and Information, University of Pittsburgh, Pittsburgh, PA, USA}{aba70@pitt.edu}{https://orcid.org/0000-0001-6406-9603}{}
\author{Balaji Palanisamy}{School of Computing and Information, University of Pittsburgh, Pittsburgh, PA, USA}{bpalan@pitt.edu}{https://orcid.org/0000-0002-0282-0913}{}
\author{Prashant Krishnamurthy}{School of Computing and Information, University of Pittsburgh, Pittsburgh, PA, USA}{prashk@pitt.edu}{https://orcid.org/0009-0004-8598-2126}{}

\authorrunning{A. Alhaidari, B. Palanisamy, and P. Krishnamurthy} 
\Copyright{Abdulrahman Alhaidari, Balaji Palanisamy, and Prashant Krishnamurthy}



\nolinenumbers

\funding{This material is based upon work supported by the National Science Foundation under Grant \#2020071. Any opinions, findings, and conclusions or recommendations expressed in this material are those of the authors and do not necessarily reflect the views of the National Science Foundation.}

\nolinenumbers

\begin{document}

\maketitle

\begin{abstract}
Billions of dollars are lost every year in DeFi platforms by transactions exploiting business logic or accounting vulnerabilities. Existing defenses focus on static code analysis, public mempool screening, attacker contract detection, or trusted off-chain monitors, none of which prevents exploits submitted through private relays or malicious contracts that execute within the same block. We present the first decentralized, fully on-chain learning framework that: (i) performs gas-prohibitive computation on Layer-2 to reduce cost, (ii) propagates verified model updates to Layer-1, and (iii) enables gas-bounded, low-latency inference inside smart contracts. A novel Proof-of-Improvement (PoIm) protocol governs the training process and verifies each decentralized micro update as a self-verifying training transaction. Updates are accepted by \textit{PoIm} only if they demonstrably improve at least one core metric (e.g., accuracy, F1-score, precision, or recall) on a public benchmark without degrading any of the other core metrics, while adversarial proposals get financially penalized through an adaptable test set for evolving threats. We develop quantization and loop-unrolling techniques that enable inference for logistic regression, SVM, MLPs, CNNs, and gated RNNs (with support for formally verified decision tree inference) within the Ethereum block gas limit, while remaining bit-exact to their off-chain counterparts, formally proven in Z3. We curate 298 unique real-world exploits (2020 - 2025) with 402 exploit transactions across eight EVM chains, collectively responsible for \$3.74 B in losses. We demonstrate that on-chain ML governed by \textit{PoIm} detects previously unseen attacks with over 97\% attack detection accuracy and 82.0\% F1. A single inference, such as one made via an external call, typically incurs zero cost. Fully on-chain inference consumes 57,603 gas ($\approx\$0.18$) for linear models, 143,647 gas ($\approx\$0.49$) for CNN(F2, K1), and 506,397 gas ($\approx\$1.77$) for CNN(F8, K4) on L1 (e.g., Ethereum). Our results show that practical and continually evolving DeFi defenses can be embedded directly in protocol logic without trusted guardians, and our solution achieves highly cost-effective protection while filling a critical gap between vulnerability scanners and real-time transaction screening.

\end{abstract}

\section{Introduction}
Vulnerabilities in decentralized finance (DeFi) protocols are triggered through transactions~\cite{auer2024technology}. Attackers usually do not bypass contract safeguards directly~\cite{sayeed2020smart}, instead they craft transactions that invoke legitimate functions to trigger state changes that the protocol did not intend. Exploits occur when contracts proceed with malicious inputs while assuming invariant state conditions, such as executing a withdrawal without checking balance or allowance or privileged operations without enforcing access control~\cite{perez2021smart}. These attacks can be single transactions or atomic sequences that satisfy syntax checks but produce unauthorized asset control. The execution logic can be formally valid, yet produce outcomes that violate the protocol's security assumptions~\cite{carter2021defi}. Vulnerabilities come from code bugs and manipulating contracts. Attackers chain operations (for example, in atomic transactions) that appear normal in unexpected ways to generate exploits~\cite{chen2020survey}. However, most previous work focuses on static code-level bugs~\cite{piantadosi2023detecting} and overlooks protocol-level flaws~\cite{zhang2023demystifying}.

Static analysis detects patterns that violate predefined coding conventions, but fails to capture logic flaws that depend on contract state, cross-function flows, or interactions across multiple protocols~\cite{chaliasos2024smart}. For instance, Cream Finance lost over \$130M through an interaction with another protocol that allowed borrowing without triggering the appropriate collateral checks~\cite{nexus2021cream}. Formal verification tools prove that certain invariants hold under all code paths but do not encode financial semantics or simulate attacker incentives. Most verification frameworks cannot model adversarially composed transaction sequences, in which attackers combine individually valid operations to produce exploits. Thus, zero-day vulnerabilities continue to appear in DeFi~\cite{zhang2023demystifying}. Tools for detecting malicious smart contracts assume a time window between contract deployment and first exploitation~\cite{ren2024lookahead}, during which vulnerabilities can be analyzed. In practice, attackers deploy and execute the exploit within a single block or bypass deployment entirely by sending malicious transactions from externally owned accounts (EOA)~\cite{li2024characterizing}.

Other methods, such as post-attack analysis, provide insight into what has happened but offer no protection. They begin only after an exploit has occurred and rely on retrospective debugging of protocol states~\cite{lam}. However, recent attacks in DeFi have compromised over \$79.8 billion in DeFi assets and only \$6.7 billion of them have been recovered~\cite{defiRektDatabase}. This emphasizes a core challenge in shared threat intelligence, particularly for aftermath attack analysis, where dissemination is usually delayed. This delay increases the likelihood of repeat attacks on existing deployed protocols that may harbor the same unaddressed vulnerabilities. Methods such as front-running protection~\cite{zhang2023your} monitor public mempools (public queue of transactions) but miss transactions submitted through private relays such as Flashbots~\cite{flashbots}. While off-chain monitoring systems (e.g.,~\cite{zhang2023your, alhaidari2025protecting, ren2024lookahead}) can detect some attacks, they typically focus on specific classes (e.g., flash loans, reentrancy). Moreover, their effectiveness is diminished by latency, which can lead to costly responses (due to gas fees) and reduced efficacy, particularly against sophisticated attacks~\cite{wang2023survey}. Therefore, there is a need for protocol-integrated security mechanisms that add an evolving layer of protection (i.e., Intrusion Prevention System (IPS)) to smart contracts and react in real time without relying on external entities, such as pre-attack (e.g., auditing) or post-attack (i.e., attack tracing) countermeasures. On the other hand, DeFi protocols currently lack mechanisms to evaluate transaction intent during execution other than hard-coded logic, which makes them vulnerable to attacks that were not anticipated during smart contract design.

Solutions driven by machine learning (ML) show strong capabilities in attack detection \cite{guo2023review, kayikci2024blockchain}. However, deploying ML models directly to Layer-1 blockchains such as Ethereum faces major obstacles in computation and storage. The computational demands of ML algorithms result in prohibitive gas costs. Ethereum's block gas limits ($\approx 30$ million units per block~\cite{ethereum}) impose constraints on the computational complexity of transactions, making complex ML models impossible to run in a single transaction. A basic model inference might consume 30-50\% of an entire block. Storage costs create another barrier. Saving model parameters on-chain is extremely expensive~\cite{kayikci2024blockchain}. These combined limitations make direct Layer-1 deployment of ML solutions economically and technically impractical for real-world deployment.

To address the above-mentioned gaps, we propose a decentralized training architecture where all model training and governance occur on Layer-2 (L2), while inference is optimized and happens on Layer-1 (L1) under strict gas constraints (Table \ref{tab:comparison-symbolic}). In our framework, an ML model is trained to provide a layer of security for smart contracts running on the Ethereum (L1) network. Our framework leverages L2 (e.g., Optimism rollup) for intensive computation and L1 for optimized inference to overcome L1 resource constraints (such as computation limits and gas fees). The rollup provides cost-effective computation while inheriting the base blockchain's security guarantees~\cite{thibault2022blockchain}. Training is performed on L2, and L1 updates are governed by L2 decentralized nodes, where computation is cheaper, and the learned model is cryptographically verified and propagated to L1 for inference. Inference runs at zero cost through read-only classifiers (for example, pure or view functions) or is fully verified on L1 embedded in on-chain contracts for real-time transaction classification. To support a range of use cases, we propose two tiers: \textit{fully zero-cost inference} (for users and smart contracts) on L1, and \textit{fully on-chain ML on L1}. The system acts as a transaction gatekeeper (i.e., firewall) even for high-throughput or low-value use cases. Our approach is model-agnostic and supports various ML algorithms, including linear algorithms (e.g., logistic regression, SVMs) and non-linear models (e.g., neural networks up to 10 layers, including 10-layer convolutional neural networks (CNN)). These models are optimized and serialized into constant-time evaluation logic fully bounded by L1 constraints. We found that even low-overhead mechanisms (linear models) are sufficient to detect a wide range of attacks. The models are trained using micro-steps by decentralized peers based on the collective knowledge of the DeFi platforms, and their formally verifiable performance matches traditional on-chain counterparts without approximation, enabling detection of both known and novel transaction behaviors.

As DeFi exploits evolve over time and to maintain continuous learning of new exploits as they appear, we introduce the notion of \textit{Proof-of-Improvement (PoIm)}. \textit{PoIm} is a decentralized protocol on L2 that governs and verifies micro-step model training and its L1-propagated updates. A deployed model is designed to be shared globally across DeFi protocols, allowing any platform to contribute by training it on one candidate (malicious or normal) transaction at a time. These DeFi platforms collectively shape a model that becomes robust over time and enables unified sharing of attack intelligence. All updates are evaluated against an on-chain (L2) committed, agreed-upon benchmark of past exploit and benign transaction data. Updates are accepted only if they demonstrably improve at least one of the key performance metrics (namely precision, recall, accuracy, or F1-score) without degrading others, as evaluated by \textit{PoIm} against the on-chain benchmark and it is verifiable by any node. Submitters are rewarded in proportion to the verified performance gain, while failed proposals lose their stake. If a malicious update seemingly improves metrics but enables a detection bypass, peers can vote to roll back the model to the last stable version. \textit{PoIm} enables the L1 classifier to evolve with newly emerging exploits without relying on centralized oversight.

For our evaluation, since there is no publicly available transaction data for DeFi attacks, we manually collected 298 confirmed exploit transactions from real DeFi attacks across eight major blockchains: \textit{Ethereum, Binance Smart Chain (BSC), Polygon, Avalanche, Arbitrum, Fantom, Moonriver, and Base}. Our exploit collection generalizes to cover attacks that exploited smart contract vulnerabilities in the past five years (2020 - April 2025), reported in news, social media platforms (e.g. X), blogs, and DeFi attack documentation. We utilize blockchain data from public explorers (for example, Etherscan~\cite{etherscan}) and Web3 Remote Procedure Calls (RPCs) (e.g. using libraries/services like Web3.js~\cite{web3js} and Alchemy~\cite{alchemy}), along with DeFiHackLabs~\cite{defihacklabs}, the DeFi Rekt Database~\cite{defiRektDatabase}, and DeFiLlama~\cite{lam} as ground truth for guiding historical exploit collection. Each transaction contains the exact exploit transaction data used by the attacker, as observed by the smart contract at execution time, including call parameters, sender addresses (EOAs or smart contracts), and relevant blockchain state at the time of the exploit. In addition, we collected comprehensive metadata for each attack, including protocol names, exploited functions, attack methods, root causes, and financial losses. We use these real-world exploits, which collectively caused over \$3.74 billion in losses. Our evaluation shows that our approach is highly gas-efficient. CNN and RNN models governed by our decentralized \textit{PoIm} protocol achieve over 97\% recall and up to an 82\% F1-score on unseen DeFi exploits.

\noindent\textbf{Contributions.} This paper makes the following contributions:
\begin{itemize}[leftmargin=*]
  \item We design a decentralized learning framework with training and governance on L2 and enable two tiers: zero-cost inference and fee-optimized inference on L1 for real-time classification of DeFi transactions.
  \item We introduce \textit{PoIm}, a decentralized L2 protocol that governs model training and propagates verified updates to L1 for inference.
  \item We formally verify inference correctness, model update integrity, and L1/L2 consistency under gas and computation constraints.
 \item We evaluated our framework on a curated set of 298 manually collected real-world exploits. It achieves high detection performance: SVM reaches an F1-score of 80\%, and CNN(F4, K4) achieves 82\% F1, 0.9004 accuracy, and over 97\% recall on unseen exploits. L1 inference is efficient, requiring only 57,603 gas for linear models and 143,647 gas for CNN(F2, K1). Zero-cost inference is supported via external EVM nodes.

\end{itemize}

\begin{table*}[t]
\small
\caption{
Comparison of our work with prior on-chain ML studies. 
\textbf{DT} = Decentralized Training, 
\textbf{MC} = Model Consistency (On-/Off-chain), 
\textbf{G} = Governance, 
\textbf{DF} = DeFi Focus, 
\textbf{IC} = Inference Cost, 
\textbf{TM} = Trust Model, 
\textbf{VM} = Validation Method.  
}
\label{tab:comparison-symbolic}
\centering
\begin{tabular}{l@{\hskip 3pt}c@{\hskip 3pt}c@{\hskip 3pt}c@{\hskip 3pt}c@{\hskip 3pt}c@{\hskip 3pt}c@{\hskip 3pt}c}
\toprule
\textbf{Study} & \textbf{DT} & \textbf{MC} & \textbf{G} & \textbf{DF} & \textbf{IC} & \textbf{TM} & \textbf{VM} \\
\midrule
\textbf{Our Work} & \cmark & \cmark & \cmark & \cmark & Zero / Low & \textit{\textit{PoIm}} & On-chain + commit-reveal \\
\textbf{ML2SC~\cite{li2024ml2sc}} & \xmark & \textit{Partial}\footnotemark[1] & \xmark & \xmark & High & \xmark & \xmark \\
\textbf{LMST~\cite{sham2025generation}} & \xmark & \textit{Partial}\footnotemark[2] & \xmark & \xmark & Mod-High & \xmark & \xmark \\
\textbf{opML~\cite{conway2024opml}} & \xmark\footnotemark[3] & \cmark & \cmark & \xmark & Low* & AnyTrust & Fraud-proof \\
\bottomrule
\end{tabular}

\vspace{0.5em}
\footnotesize
\textsuperscript{1} Minor mismatches due to PRBMath; 
\textsuperscript{2} Accuracy drop from fixed-point; 
\textsuperscript{3} Training off-chain, validated via fraud proof; 
* Optimistic assumption, only $O(1)$ on-chain arbitration.
\end{table*}

\section{Background}\label{sec:background}

\textbf{EVM blockchains.}  
Ethereum is the largest Ethereum Virtual Machine (EVM) blockchain. It runs with a consensus mechanism as a peer-to-peer network that supports programmable logic or smart contracts executed by the EVM~\cite{ethereum}. There are two main types of accounts: Externally controlled accounts (EOAs), operated by cryptographic keys, and smart contract accounts (SCAs), which execute embedded logic upon invocation~\cite{saraf2018blockchain}. EOAs initiate transactions, while SCAs automate protocol behavior or malicious activity~\cite{sayeed2020smart}. \\ 
\textbf{Layer 2 blockchains.} Layer 2 (L2) blockchains (e.g., Optimism, Arbitrum, Base) are built on top of Layer 1 (L1) platforms (e.g., Ethereum, Binance Smart Chain (BSC)) to improve scalability and reduce costs~\cite{sguanci2021layer}. In contrast to L1s that execute and store every transaction directly on-chain~\cite{ethereum}, L2s batch many transactions and periodically submit compressed proofs to the L1~\cite{xu2022l2chain}. This design allows L2s to offer significantly lower gas fees and faster execution and inherits the security guarantees from L1~\cite{sguanci2021layer}. Notably, L2s support the same smart contract logic as L1s but with relaxed resource constraints, which makes them ideal for computationally heavy tasks like model training or repeated inference.

\noindent\textbf{Transactions.}  
Transactions are the main component used to facilitate both asset movement and protocol interactions. A nonce in a transaction acts as a sequence counter to prevent replay attacks, where the attacker could replicate the transaction. The gas fee, influenced by user-specified price and the complexity of transactions, determines inclusion priority, while the gas limit defines the upper computational allowance per transaction~\cite{wood2014ethereum}.\\
\noindent\textbf{Exploit transactions.} Most of the DeFi attacks succeed through a maliciously crafted sequence of instructions (e.g., internal transactions)~\cite{dwivedi2024novel}. These attacks exploit weaknesses in smart contracts rather than modifying contract code at the infrastructure level~\cite{zhou2023sok}. Attackers manipulate transaction parameters, call sequences, and attempt to manipulate permission states through public interfaces to gain unauthorized assets~\cite{sayeed2020smart}. \\
\noindent\textbf{Mempool and private relays.}  
Blockchain transactions are submitted either through the public transaction buffer, known as the mempool, or via private relays~\cite{flashbots}, where the transaction is sent directly to miners. Transactions submitted via private relays remain hidden from the public, whereas those using the mempool are broadcast and await confirmation~\cite{choudhuri2024mempool}. Nodes share this queue across the network. The selection of which transactions to include in a block is typically based on miners/validators' incentives and the gas fees offered. Transactions with higher fees are generally prioritized~\cite{liu2022empirical}. The mempool is publicly visible, enabling a brief window during which real-time monitoring can be used to detect malicious behavior, such as front-running~\cite{heimbach2022eliminating}. Since each node maintains a synchronized copy of unconfirmed transactions, this window supports early threat detection~\cite{gervais2016security}. However, detection is not always possible due to the complexity of some attacks, and in many cases, the malicious transaction is submitted directly to miners/validators, bypassing public visibility~\cite{messias2023dissecting}.\\
\textbf{DeFi and smart contract vulnerabilities.}
Decentralized finance (DeFi) is blockchain-based financial services operating without centralized intermediaries~\cite{chen2020blockchain}. Smart contracts, self-executing programs deployed on blockchain platforms such as Ethereum, form the backbone of DeFi platforms~\cite{ethereum}. These contracts facilitate autonomous, trust-minimized interactions, enabling services such as decentralized exchanges (DEXs), lending, asset management, and stablecoins. The complexity and transparency of smart contracts introduce significant risks. Code-level vulnerabilities, flawed logic, and unintended transaction sequences can lead to severe security breaches and substantial financial loss~\cite{haouari2024vulnerabilities}. Common exploit classes include reentrancy attacks, access control failures, approval abuses, and flash loan-based manipulations~\cite{xu2022reap}. Exploits often occur rapidly and irreversibly, exploiting the immutable nature of blockchain transactions. Although these exploit classes are studied, there is a need for defense solutions that are capable of mitigating malicious transaction behaviors in real time for diverse exploit types~\cite{dwivedi2024novel}.

\subsection{Limitations of Existing Defenses}
This section discusses fundamental limitations of existing preventive measures and real-time defense mechanisms. First, off-chain detection systems such as LookAhead~\cite{ren2024lookahead} rely on a temporal gap between transaction submission and execution. When an attacker deploys a malicious contract and immediately initiates an exploit within a single block, these systems fail to respond on time. Second, private relay services like Flashbots~\cite{flashbots} allow attackers to bypass mempool-based detection by submitting transactions directly to miners/validators. In such settings, traditional monitoring tools lose all visibility, allowing stealthy attacks without external traceability.

Third, signature-based detection methods depend on identifying known function selectors or matching call patterns against static signatures~\cite{alhaidari2025protecting}. This approach fails when attackers use proxy contracts, delegate calls, or obfuscated logic flows, where surface-level transaction signatures are intentionally masked. Fourth, current on-chain defenses lack the ability to validate transaction behavior dynamically during execution. Existing systems either rely on static pre-deployment audits or basic access control checks, leaving dynamic runtime behavior, such as unauthorized fund movements or privilege escalations, undetected.

Fifth, most prior defenses prioritize contract vulnerabilities rather than transaction behavior. However, even with this focus, a recent study~\cite{chaliasos2024smart} found that automated tools can detect only 8\% of vulnerabilities. Another study found that existing tools can detect only 20\% of vulnerabilities~\cite{zhang2023demystifying}, which leaves 80\% undetected. Consequently, they can be exploited and cause huge financial losses. One reason is that not all bugs are exploitable, and not all correct logic code is vulnerability-free~\cite{perez2021smart}. Thus, the focus on code logic overlooks the fact that many exploits occur through crafted transaction sequences exploiting valid contract interfaces without exploiting traditional code vulnerabilities. 

Also, existing solutions typically adopt a centralized approach. Systems that depend on trusted off-chain detectors or external alert mechanisms inherently introduce single points of failure and trust dependencies incompatible with DeFi principles.

Finally, another category of defense focuses on pre-interaction analysis to identify inconsistencies between a project's documentation and its on-chain bytecode. DeFiAligner~\cite{gan2024defialigner}, for instance, uses symbolic execution and Large Language Models to detect when the implementation and documented logic of a smart contract do not match. This approach is valuable for auditing projects and protecting users from being misled by inaccurate documentation. However, it does not provide real-time defense against malicious transaction behavior when it is executed. It also cannot detect a novel exploit sequence that uses valid functions in an unexpected (malicious) way, which is the gap our work addresses.

\subsection{Limitations of On-Chain ML} 
Embedding machine learning models into L1 smart contracts is resource-intensive, and replicating off-chain models directly on-chain can be impossible. L1 blockchains are not designed for heavy computing~\cite{abed2021analysis}. However, inference, if designed without inheriting the complexity of off-chain models (e.g., using TensorFlow~\cite{abadi2016tensorflow}, Keras~\cite{ketkar2017introduction}, Scikit-learn~\cite{kramer2016scikit}) yet performing equivalently, can be feasible if heavy computations are separated from inference. 

Previous designs (e.g.~\cite{li2024ml2sc, sham2025generation}) face many practical barriers, including prohibitive gas costs for inference, inconsistent behavior between off-chain and on-chain models due to numerical limitations, and incompatibility with smart contract languages lacking native support for floating-point operations, which causes model output deviation (off-chain vs on-chain counterpart). For example, the authors in~\cite{sham2025generation} reported inconsistent accuracy between the PyTorch and on-chain models: 86.00\% off-chain versus 81.00\% on-chain. The deployment cost also exceeds the current Ethereum block gas limit, with a reported cost of 73,721,648 gas. This is nearly twice Ethereum's current block gas limit, which is approximately 30 million gas units.

Our work achieves exact consistency (formally and empirically verified) between off-chain and on-chain model outputs. Unlike prior methods, our on-chain design uses an optimized smart contract that replicates the off-chain model's behavior without approximation to achieve the same outcome. This will allow DeFi attacks mitigation dynamically using practical, low-cost, and fully decentralized on-chain machine learning.

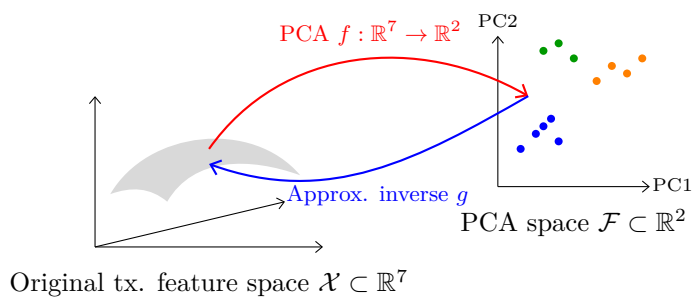
\begin{figure}[t]
\centering
\begin{tikzpicture}

\draw[->] (0, 0) -- ++(0, 2);
\draw[->] (0, 0) -- ++(2.5, 0.6);
\draw[->] (0, 0) -- ++(3, 0) node[midway, below, yshift=-0.5em]
    {Original tx. feature space ${\cal X} \subset \mathbb{R}^7$};

\draw[fill=gray!30, draw=none, shift={(0.2, 0.7)},scale=0.5]
  (0, 0) to[out=20, in=140] (1.5, -0.2) to [out=60, in=160]
  (5, 0.5) to[out=130, in=60] cycle;

\draw[->] (5.3, 0.8) -- ++(0, 2);
\draw[->] (5.3, 0.8) -- ++(2, 0) node[midway, below, yshift=-0.5em]
      {PCA space ${\cal F} \subset \mathbb{R}^2$};

\foreach \x/\y in {5.6/1.3, 5.8/1.5, 6.0/1.7, 6.1/1.4, 5.9/1.6} {
    \fill[blue] (\x,\y) circle (1.5pt);
}
\foreach \x/\y in {6.6/2.2, 6.8/2.4, 7.0/2.3, 7.2/2.5} {
    \fill[orange] (\x,\y) circle (1.5pt);
}
\foreach \x/\y in {5.9/2.6, 6.1/2.7, 6.3/2.5} {
    \fill[green!60!black] (\x,\y) circle (1.5pt);
}

\node at (7.6, 0.83) {\scriptsize PC1};
\node at (5.3,3) {\scriptsize PC2};

\draw[thick,->,red]
  (1.5, 1.3) to [out=55, in=150] node[midway, above, xshift=6pt, yshift=2pt]
  {\footnotesize PCA $f: \mathbb{R}^7 \to \mathbb{R}^2$} (5.7, 2);

\draw[thick,->,blue]
  (5.7, 2) to [out=-150, in=-20] node[midway, below, xshift=2pt, yshift=-2pt]
  {\footnotesize Approx. inverse $g$} (1.5, 1.1);

\end{tikzpicture}
\caption{PCA reduces 7D attack transaction features from the original feature space ${\cal X} \subset \mathbb{R}^7$ to a 2D latent space ${\cal F} \subset \mathbb{R}^2$ for pattern clustering.}

\label{fig:pca_projection}
\end{figure}

\section{DeFi Attacks and Threat Model}

The current state of defenses is not sufficient to mitigate the growing threat of unauthorized fund exploits in Decentralized Finance (DeFi). Recent studies~\cite{chaliasos2024smart,li2024defitail,choudhuri2024mempool,zhou2023sok,zhang2023demystifying} show that attack techniques are rapidly advancing but defenses remain inadequate. To address these threats, we need a new defense model based on a real-time, in-protocol (smart contract-level) transaction classification system that must operate independently of external monitors or pre-defined signature lists~\cite{zhou2023sok}, and it must work even when transactions are hidden from the public mempool. It should also generalize to both known and novel (zero-day) attacks that exploit DeFi contract flaws. However, statistical methods have shown significant promise in detecting not only known attacks but also previously unseen zero-day exploits~\cite{guo2023review}. This naturally raises several key questions: Can data from past DeFi attacks help us predict and prevent future ones? What are the possible designs for trustless, cost-effective, on-chain inference systems that can evolve with the attack landscape and be governed by decentralized peers? To answer these questions, we investigate whether real historical exploits exhibit detectable patterns that statistical techniques can reliably uncover and learn from.

\begin{figure}[h]
    \centering
    \includegraphics[width=0.8\linewidth]{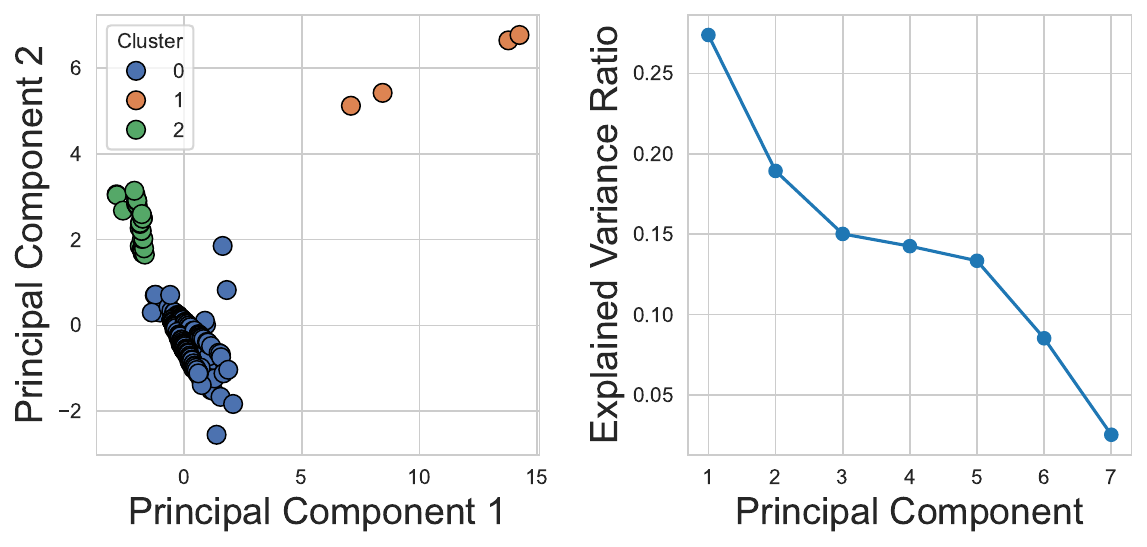}
\caption{Left: PCA projection of DeFi attack transactions using two principal components with KMeans clustering ($k=3$). Right: The scree plot shows the explained variance ratio per principal component. Even though PC1 and PC2 capture less than 50\% of the total variance, the projection shows distinct DeFi attack transaction patterns.}
    \label{fig:PCA}

\end{figure}

\subsection{Empirical Patterns of DeFi Attacks}

To explore the feasibility of detecting DeFi attacks using only transaction data observable by smart contracts, we analyzed 402 confirmed DeFi exploit transactions (for 298 attacks). Transaction data includes complete EVM-level execution context such as gas usage, transferred value, calldata, and block metadata. Since previous studies offer limited analysis of DeFi attacks from a transactional perspective, we investigate whether exploit transactions exhibit measurable similarities that support transaction-level detection.

\begin{table}[h]
\centering
\small
\caption{Transaction Metadata Visible to Smart Contracts During Execution}
\begin{tabular}{ll}
\hline
\textbf{Attribute} & \textbf{Description} \\
\hline
\texttt{msg.sender} & Sender address (initiator of the call) \\
\texttt{msg.value} & ETH amount sent with the call \\
\texttt{msg.data} & Calldata containing function selector and arguments \\
\texttt{tx.origin} & Original externally-owned sender of the transaction \\
\texttt{gas} & Remaining gas (via \texttt{gasleft()}) \\
\texttt{gas\_price} & Effective gas price paid (via \texttt{tx.gasprice}) \\
\texttt{msg.to} & Receiving contract's address (via \texttt{address(this)}) \\
\hline
\end{tabular}
\label{tab:visible_tx_metadata}
\end{table}

We selected the numerical features observable during smart contract execution (Table \ref{tab:visible_tx_metadata}) and excluded transaction hashes or addresses. Features were standardized, and dimensionality was reduced (Figure~\ref{fig:pca_projection}) using Principal Component Analysis (PCA). We applied KMeans clustering with $k=3$ to detect grouping behavior across attack transactions.

The PCA projection preserved 46.3\% of the variance in two dimensions (Figure~\ref{fig:PCA}). Out of 402 exploits, 355 (88.3\%) formed a distinct cluster. This suggests high behavioral consistency across transactions despite protocol and chain differences. Clustering quality was supported by a Calinski-Harabasz index of 981.78 and a Silhouette score of 0.773, both indicating well-separated, compact clusters.

These findings indicate that DeFi attacks share consistent runtime characteristics. The ability of PCA + KMeans to cluster these attacks supports the feasibility of transaction-level classification during execution. We classify the root causes of these attacks into five empirically supported categories, such as access control failures and business logic flaws (see Table \ref{RC} in Appendix).

\noindent\textbf{Exploit Execution Patterns.}
Despite different root causes, several runtime-level patterns are consistent across attacks: 
\begin{itemize}
    \item \textbf{Atomicity:} 71.2\% of exploits execute within a single transaction with no prior on-chain activity.

    \item \textbf{Benign interface misuse:} 67.5\% invoke functions like \texttt{withdraw()}, \texttt{approve()}, or \texttt{sweep()} with malicious arguments.

    \item \textbf{Cross-chain replication:} Identical bytecode deployments (9.7\%) are exploited across chains~\cite{he2020characterizing} (e.g., Hedgey \cite{hedgey}).
\end{itemize}
These patterns further support the view that exploitability is determined by transaction behavior rather than only static code.

\subsection{Challenges in Real-Time Defense}
\label{sec:defi-challenges}
We identified several key challenges that prevent current systems from defending against these attacks in real time:
\begin{itemize} 
    \item[\textit{\textbf{C1.}}] \textbf{Private relay invisibility:} Transactions sent through Flashbots~\cite{flashbots} and similar relays bypass the public mempool, evading pre-inclusion detection.
    \item[\textit{\textbf{C2.}}] \textbf{Multiple attack paths per contract:} A single protocol may contain unrelated flaws, making static patching or signature detection ineffective.
    \item[\textit{\textbf{C3.}}] \textbf{Benign-looking interfaces:} Safe-looking functions are abused with crafted inputs, undermining static signature-based detection.
    \item[\textit{\textbf{C4.}}] \textbf{Cross-chain propagation:} An undetected exploit on one chain quickly propagates to other protocols with shared logic.
    \item[\textit{\textbf{C5.}}] \textbf{No rollback:} Once executed, DeFi transactions are final. Most protocols do not have pause switches or delayed execution.
\end{itemize}
 
\subsection{Threat Model and Assumptions}
Our system defends against two primary threat scenarios.
\ding{172} An attacker crafts a malicious transaction targeting a vulnerable contract. The transaction is evaluated by our on-chain ML classifier before execution. The attacker may vary calldata, timing, or submission channel (e.g., private relay), but cannot alter the deployed model or governance system. \ding{173} A malicious peer attempts to poison the training process by submitting manipulated updates that degrade model performance. The system detects and rejects such updates via on-chain benchmarking.
Insider threats, compromised signing keys, and off-chain infrastructure attacks are out of scope for this threat model.
\vspace{1mm}
\begin{figure}[ht]
\centering
\includegraphics[width=0.75\linewidth]{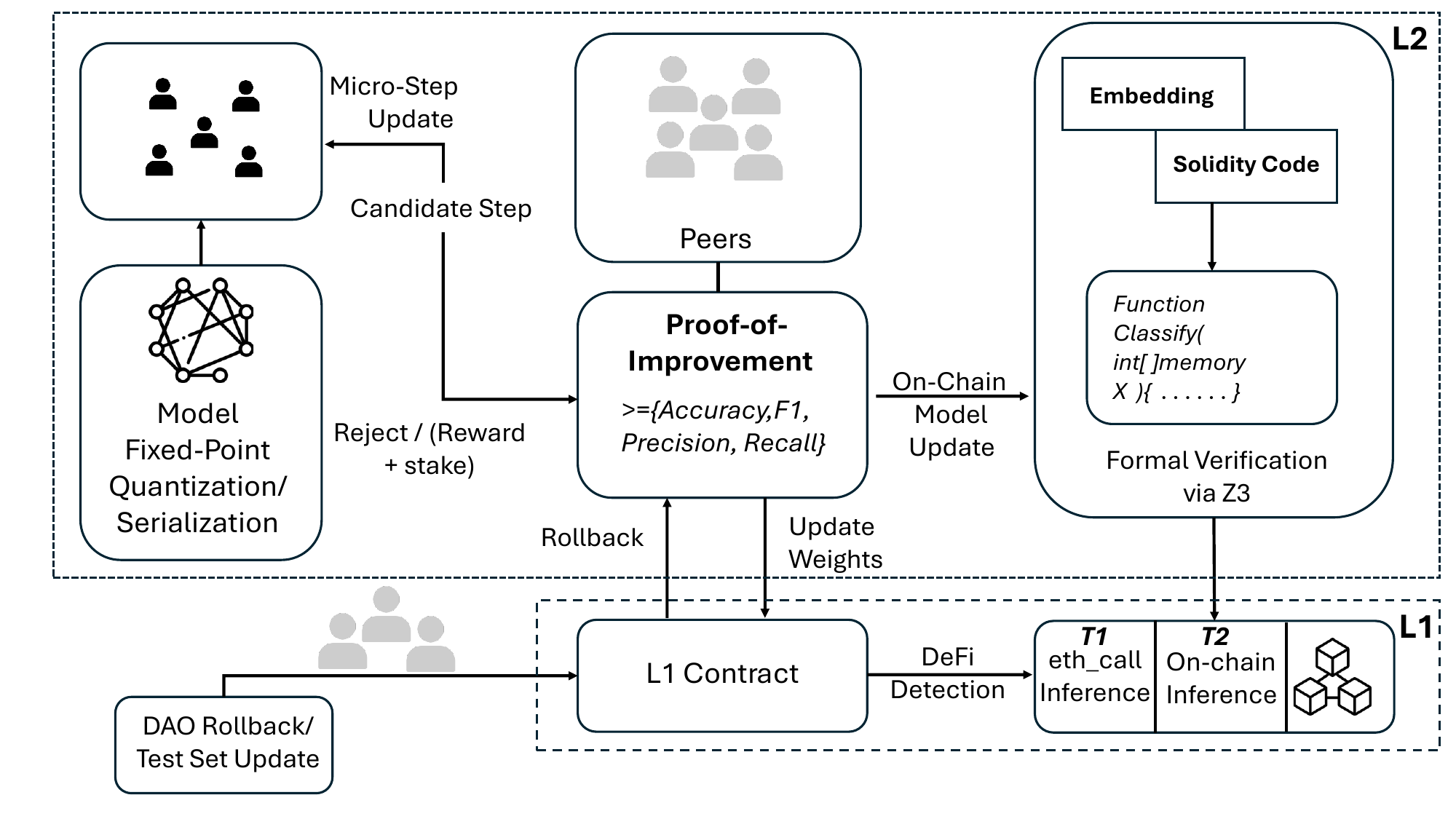}
\caption{Framework overview.}
\label{fig:zero_cost_inference}
\end{figure}
\section{Decentralized Training, Inference, and Governance Framework} 
Our objective is to integrate decentralized learning with verifiable and efficient on-chain detection into the execution paths of DeFi contracts to mitigate exploit transactions in real time. We address limitations in existing works (e.g.,~\cite{li2024ml2sc, zhang2023your, alhaidari2024flashguard, ren2024lookahead, sham2025generation}) from two perspectives: DeFi attack mitigation and on-chain ML design. First, we detect private relay attack transactions and generalize the detection mechanism to unseen attacks in real time. Second, we ensure bitwise consistency between off-chain and on-chain models, and cost-effective decentralized training and inference. We also guarantee bounded resource consumption, such as gas fees for training and inference. Furthermore, we introduce a decentralized mechanism that enables peers (i.e., DeFi platforms) to collectively train ML/DL models (e.g., for attack defense) by proposing training samples. These proposed updates are transparently governed, validated, and challengeable by others. An overview of our end-to-end framework is presented in Figure~\ref{fig:zero_cost_inference}.

\begin{figure}[ht]
\centering
\begin{tikzpicture}[
    font=\scriptsize,
    neuron/.style={circle, draw=black, fill=white, minimum size=0.6cm, inner sep=1pt, thick},
    arr/.style={-Latex, thick},
    box/.style={draw=black, fill=gray!10, minimum height=0.9cm, minimum width=4.5cm, align=center, font=\scriptsize},
    label/.style={font=\sffamily\scriptsize\bfseries},
  ]

  \node[label] at (1.2, 2.2) {Quantized Model};

  \node[neuron] (in0) at (0, 1.5) {$h^+_{t\,0}$};
  \node[neuron] (in1) at (0, 0.7) {$h^+_{t\,1}$};
  \node[neuron] (in2) at (0, 0) {$\vdots$};
  \node[neuron] (in3) at (0, -0.9) {$h^+_{t\,n}$};

  \node[neuron] (out0) at (2.1, 1.5) {$b^{(2)}$};
  \node[neuron] (out1) at (2.1, 0.7) {$b^{(2)}$};
  \node[neuron] (out2) at (2.1, 0) {$\vdots$};
  \node[neuron] (out3) at (2.1, -0.9) {$b^{(2)}$};

  \foreach \i in {0,1,3}
    \foreach \j in {0,1,3}
      \draw[arr] (in\i) -- (out\j);

  \draw[arr] (2.8, 0.3) -- ++(0.7, 0);

  \node[box, right=3.7cm of in2] (serialize)
    {$[ h^\prime W_{0,1}, \mathcal{W}_{0,2}, b^{(1)}_1, \hat{\mathcal{W}}^{(2)}, \mathcal{W}_{(2)}, b^{(2)} ]$};

  \node[label, above=0.15cm of serialize] {Serialization};

\end{tikzpicture}
\vspace{2mm}
\caption{Serialization example of quantized model parameters.}
\label{fig:quantized-serialization}
\end{figure}
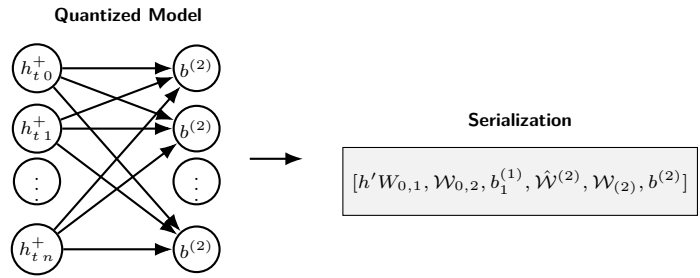

Our framework enables decentralized learning by translating traditional off-chain machine learning and deep learning model architectures into gas-efficient, formally verified Solidity contracts for blockchain execution via L2 rollups (e.g., Optimism \cite{optimism2025}). It supports a wide range of architectures, from simple to complex, and produces tamper-proof ML/DL contracts with quantized parameters serialized for on-chain weight propagation (Figure~\ref{fig:quantized-serialization}). This capability not only enhances DeFi threat detection but also provides a foundation for developing adaptive and secure financial systems in other domains. Our approach addresses the limitations of prior work discussed in Section 2.2. Previous efforts (e.g.,~\cite{li2024ml2sc, sham2025generation}) are impractical and incur high gas costs when deploying complex models on L1, which can be very complex.

In contrast, our design supports full-scale, real-world models without compromising accuracy due to Solidity's floating-point limitations, and it enables fully decentralized training and model updates. To achieve this, we introduce Proof-of-Improvement (PoIm), a decentralized governance protocol that tracks each training step and governs model updates. It allows anyone (e.g., any DeFi platform) to incrementally train the model, effectively proposing updates that improve detection metrics. Updates are predictably propagated from L2, where computation and verification are performed, to L1, where inference takes place. Peers can collectively override a propagated model update on L1 in cases of suspected malicious training (poisoning) by another peer. Overall, our decentralized design is motivated by DeFi security requirements, and we show that a decentralized, dynamic, and cost-effective ML/DL-based defense is feasible under current blockchain architectures enabled by L2, capable of mitigating attacks that have caused billions in financial losses.

\vspace{-2mm}

\subsection{Decentralized Micro-Step Training and Model Evolution}

Our system enables verifiable, decentralized training by decomposing the learning process into micro-step updates. Each micro-step corresponds to a single incremental improvement, such as retraining on one example or applying a localized adjustment, and is proposed directly on-chain. All proposals are immediately evaluated using a canonical, public test set stored on-chain.

Only updates that improve at least one core metric (accuracy, precision, recall, or F1-score) without degrading any others are accepted. This logic is enforced deterministically by the \textit{PoIm} contract, which governs model evolution and ensures that every change is auditable, trustless, and resistant to adversarial manipulation.

Every accepted or rejected update, along with its evaluation results, is permanently logged on-chain. This guarantees transparent, tamper-proof model provenance and allows peers to track, audit, or challenge any step of the training process. In contrast to centralized retraining, this decentralized micro-step protocol allows the model to evolve continuously and securely, driven entirely by peer contributions and verified in real time on-chain.

\subsection{Inference Architecture}

In our architecture, we support two cost-effective tiers for executing ML/DL inference over blockchain networks. Each tier provides a different trade-off between execution cost, verifiability, and decentralization.\\
\noindent\textbf{On-chain Logic, Off-chain Execution (zero-cost).}
In this tier, the model parameters and execution logic are stored fully on-chain. However, the actual inference computation is performed off-chain by calling \texttt{view} functions, which are executed by any EVM client. This tier incurs zero gas cost for DeFi users or protocols while ensuring that the decision logic is derived directly from verifiable on-chain bytecode and model state. Since every node executes the same bytecode deterministically, outcomes are consistent and tamper-resistant. This design is ideal for platforms seeking lightweight classification with full code transparency and no execution fees.\\
\noindent\textbf{Fully On-chain Inference (inference verifiable on-chain).}
Here, the inference is executed entirely on-chain as part of a state-modifying transaction, typically one that interacts with a DeFi protocol. The input is passed to the smart contract, which executes the classifier internally and enforces the classification result. This enables end-to-end verifiability and enforces decisions during transaction execution (e.g., rejecting or allowing access to protocol funds). While this incurs gas costs, our optimized design allows even moderately complex models, such as 10-layer CNNs with quantized integer arithmetic, to execute efficiently within Ethereum's gas limits. This tier is suited for scenarios like high TVL or security-critical DeFi functions, where classification must be verified on-chain without relying on off-chain entities to interpret results, as the outcome is enforced by on-chain consensus.
\subsection{Layer-2 to Layer-1 Computation Separation}
Model parameters are updated through decentralized training and governed by the Proof-of-Improvement (PoIm) protocol on Layer-2 (L2), where gas costs are significantly lower. These L2-validated model parameters must then be securely and accurately propagated to Layer 1 (L1) for use by inference contracts in either Tier 1 or Tier 2. To ensure that L1 inference contract parameters do not get tampered, we employ a commit-verify propagation mechanism.\\
\noindent\textbf{Commit to Model Hash (L2).} Upon acceptance of a training update (yielding new parameters $\theta'$) by the \textit{PoIm} protocol on L2, the contract computes a cryptographic hash \texttt{modelHash = keccak256(abi.encodePacked($\theta'$))}. This hash serves as a tamper-proof commitment to the new model. \\
\noindent\textbf{Transmit Commitment (L2 $\rightarrow$ L1).} The L2 contract sends this \texttt{modelHash} to L1 via the native L2-to-L1 bridge (e.g., Optimism’s \texttt{L2ToL1MessagePasser}). The hash is recorded by the L1 contract that manages model updates.\\
\noindent\textbf{Transmit Parameters and Verify (L2 $\rightarrow$ L1).} In a subsequent transaction, the full model parameters $\theta'$ are transmitted. The L1 contract recomputes the hash and verifies it against the prior commitment. A mismatch results in the rejection of the update \cite{teutsch2024scalable}.

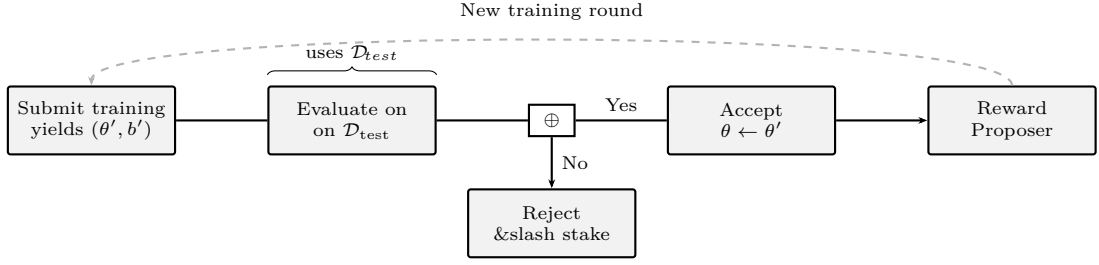
\begin{figure}[t]
\centering
\begin{tikzpicture}[
  node distance=1cm and 1.2cm,
  every node/.style={font=\scriptsize, align=center},
  box/.style={
    draw, thick, rounded corners=1pt,
    fill=gray!10,
    minimum width=2.2cm, minimum height=0.9cm
  },
  arrow/.style={thick, -{Stealth[length=1.2mm]}},
  dashedloop/.style={thick, dashed, draw=gray!60, -{Stealth[length=1.2mm]}}
]

  \node[box]   (C) {Submit training\\ yields \((\theta', b')\)};
  \node[box, right=of C] (R) {Evaluate on\\on~$\mathcal{D}_{\text{test}}$};
  \node[draw, thick, right=of R, minimum width=0.6cm] (X) {$\oplus$};
  \node[box, right=of X] (A) {Accept\\$\theta\leftarrow\theta'$};
  \node[box, right=of A] (W) {Reward\\Proposer};

  \draw[arrow] (C) -- (R) -- (X) -- node[above]{Yes} (A) -- (W);
  \draw[arrow] (X) -- node[right]{No} +(0,-0.9) 
    node[box, minimum width=2.2cm, below]{Reject\\ \&slash stake};

  \draw[decorate, decoration={brace, amplitude=3pt, raise=4pt}]
    (R.north west) -- (R.north east)
    node[midway, above=6pt] {\scriptsize uses~$\mathcal{D}_{\!test}$};

  \draw[dashedloop]
    (W.north) 
      .. controls +(0,0.8) and +(0,0.8) .. 
      node[midway, above=4pt] {\scriptsize New training round} 
    (C.north);

\end{tikzpicture}
\caption{\textit{PoIm} protocol update flow. Each proposed model is evaluated on $\mathcal{D}_{\text{test}}$ and accepted only if it improves performance.}
\label{fig:poim_flow}
\end{figure}
\vspace{-3mm}
\subsection{Formal Bit-Exact Verification}

Weights are scaled by \(S\in[10^6,10^{18}]\) then packed into \texttt{int32[]} and up-cast to \texttt{int128} where safe. A fully-connected layer \(l\) executes
$z^{(l)} = \mathrm{idiv}\!\bigl(\hat W^{(l)}z^{(l-1)}+\hat b^{(l)},\,S\bigr)$
If \(\gamma>\Delta\), sign consistency holds for all validation inputs.
For every compiled model, we prove  
$\forall x\in\mathbb Z^d:\; \mathcal F_{\text{on}}(x)=\mathcal F_{\text{off}}(x)$
under fixed-point scale \(S\). We encode both paths as bit-vector (e.g., 256-bit) formulas and ask Z3 for \texttt{expr\_on} \(\neq\) \texttt{expr\_off}. All models (linear, CNN, RNN) return \texttt{unsat}, giving a machine-checked guarantee of equality. This design-time formal proof offers a strong guarantee that the compiled on-chain model faithfully implements the intended off-chain model logic under the specified fixed-point representation for all possible inputs. It ensures the intrinsic correctness of the model's translation to Solidity. This is distinct from, yet complementary to, the operational consistency checks performed during the lifecycle of the model on-chain. Listing~\ref{lst:forward_pass} provides an example of a forward pass implemented in Solidity using fixed-point arithmetic.

\begin{lstlisting}[language=Go, caption={Forward pass example for layer $l$ using fixed-point arithmetic.},  label={lst:forward_pass}]
for (uint i = 0; i < d_l; i++) {
    z[i] = bias_l[i];
    for (uint j = 0; j < d_{l-1}; j++) {
        z[i] += idiv(weights_l[i * d_{l-1} + j] * input[j], SCALE);
    }
}
\end{lstlisting}

Beyond the formal verification of the model logic itself, our protocol incorporates runtime consistency checks at critical junctures, such as model updates on L2, propagation to L1, and sample inferences, to ensure operational integrity in runtime.

\subsection{Gas Cost and Runtime Bound Analysis}
\label{sec:gas-model}

\noindent\textbf{Opcode Budget.}
For every multiply-accumulate (MAC)
\(a \leftarrow a + \frac{w\cdot x}{S}\)
inside \texttt{classify}, the EVM executes  
(1) \texttt{SLOAD}(\(w\)) = \(G_S = 100\) gas (warm),  
(2) \texttt{CALLDATALOAD}(\(x\)) = \(G_C = 3\) gas,  
(3) \texttt{MUL} = \(G_M = 5\) gas,  
(4) \texttt{DIV} = \(G_D = 5\) gas,  
(5) \texttt{ADD} = \(G_A = 3\) gas,  
(6) loop/stack bookkeeping \(\approx G_L = 8\) gas.\footnote{Berlin fee schedule~\cite{eips}.}

Hence, the cost per MAC is  
\(g_{\text{MAC}} = G_S + G_C + G_M + G_D + G_A + G_L = 124\) gas.  
ReLU adds \(G_R = 5\) gas per activation, and bias initialization costs \(G_S + G_A = 103\) gas.\\
\noindent\textbf{Linear Classifiers (LR / SVM).}
A linear model with $d$ inputs executes one MAC per feature and adds a bias.
Hence,  
\(\boxed{G_{\mathrm{LIN}}(d)=124\,d+103}\).

For logistic regression, we approximate $\sigma(z)$ by a threshold on the logit,
therefore no extra exponentiation is incurred; the cost matches SVM. With $d{=}3$ (example feature set) the bound gives 
\(G_{\mathrm{LIN}}(3)=124\times3+103=475\ \text{gas}\), which is three orders of magnitude below the deep models and negligible at call-sites.

\noindent\textbf{CNN Bound.}
For an input of length \(d\), kernel size \(K\), and \(F\) filters,  
the convolution yields \(o = d - K + 1\) positions and evaluates \(F\,o\,K\) MACs;  
the fully connected read-out contributes a further \(F\,o\) MACs.  
Thus,  
\(\boxed{G_{\mathrm{CNN}}(d,K,F) = g_{\text{MAC}}\,F\,o\,(K+1) + G_R\,F\,o + 103\,F}\)  (1).

\noindent\textbf{RNN Bound.}
Let \(U\) be the hidden-state size, \(T\) the number of time steps, and \(d_{\text{in}} = d/T\) the per-step input width.  
A gated update performs \(U(d_{\text{in}} + U)\) MAC operations per time step. The total cost is
\(\boxed{G_{\mathrm{RNN}}(d,U,T) = g_{\text{MAC}}\,T\,U(d_{\text{in}} + U + 1) + 15\,T\,U}\) (2). \vspace{0.7mm}
\noindent We fix the total input dimensionality \(d = 3\) for both architectures. Table~\ref{fig:cnn_rnn_bounds} shows the resulting gas bounds for several example CNN and RNN configurations.

\begin{table}[h]
\centering
\small
  \renewcommand{\arraystretch}{0.88}
\begin{minipage}[t]{0.5\linewidth}
\centering
\begin{tabular}{@{}lccc@{}}
\toprule
Model & \(F\) & \(K\) & Bound \\ \midrule
CNN$_{2\times2}$ & 2 & 2 & 1\,714 \\
CNN$_{4\times2}$ & 4 & 2 & 3\,428 \\
CNN$_{8\times3}$ & 8 & 3 & 4\,832 \\ \bottomrule
\end{tabular}
\\ \vspace{0.7mm}(1)
\end{minipage}
\hfill
\begin{minipage}[t]{0.4\linewidth}
\centering
\begin{tabular}{@{}lccc@{}}
\toprule
Model & \(U\) & \(T\) & Bound \\ \midrule
RNN$_{4\times2}$ & 4 & 2 & 7\,064 \\
RNN$_{8\times4}$ & 8 & 4 & 40\,160 \\ \bottomrule
\end{tabular}
\\ \vspace{2.2mm}(2)
\end{minipage}
\caption{Analytic MAC bounds for CNN and RNN models with total input dimension \(d = 3\).}
\label{fig:cnn_rnn_bounds}
\end{table}

\begin{figure}[t]
\centering
\includegraphics[width=0.75\linewidth]{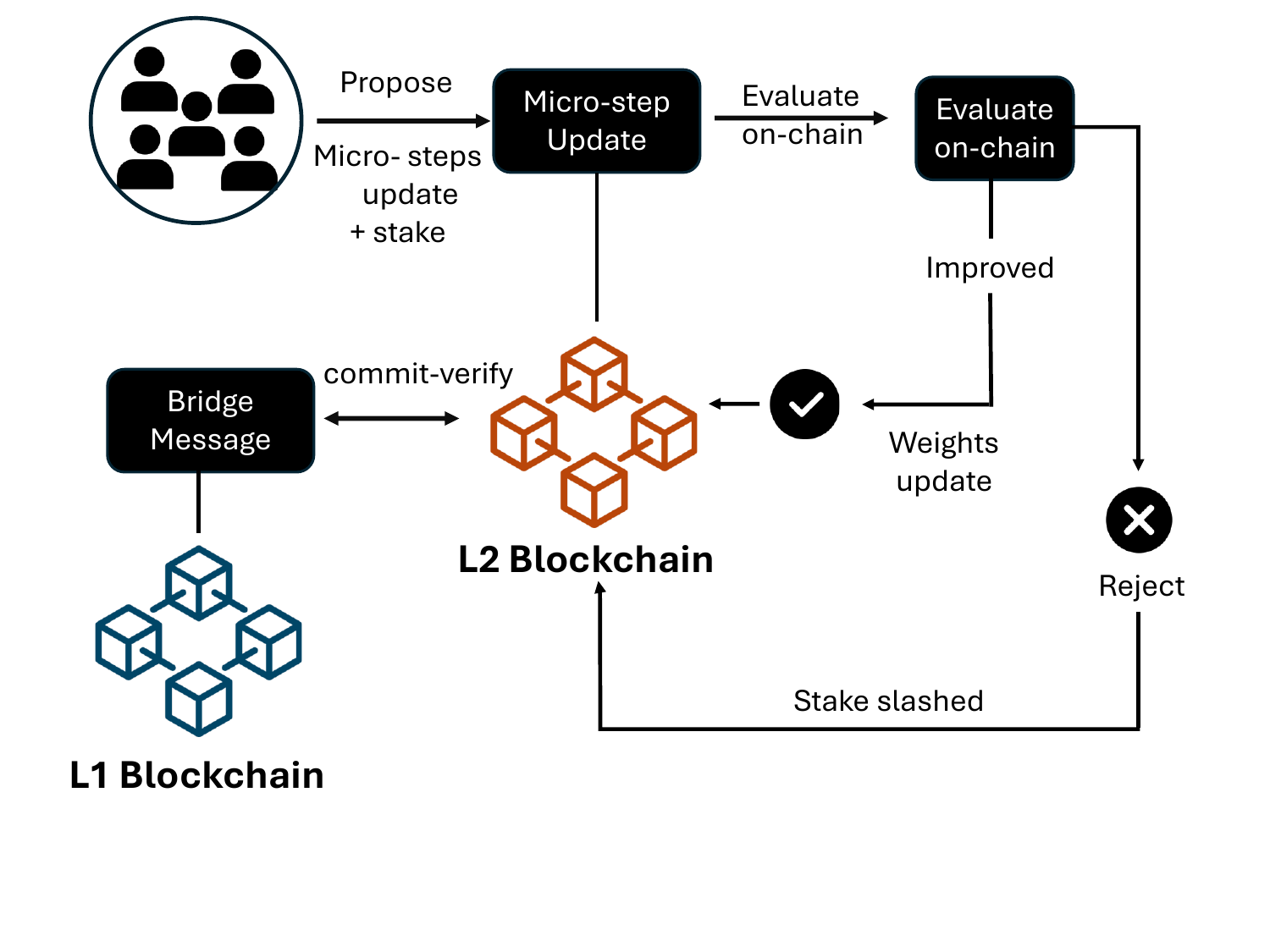}
\vspace{-9mm}

\caption{\textit{PoIm} overview.}
\label{fig:PoIm_overview}
\end{figure}

\vspace{-1mm}
\subsection{Decentralized Model Update} Only models that show improvement in multiple evaluation metrics are accepted. No external oracle trust is required and all updates are verifiable on-chain within strict gas constraints.

\begin{table}[h]
  \caption{Notations used.}
  \centering
  \small
  \begin{tabular}{@{}ll@{}}
    \toprule
    \textbf{Symbol} & \textbf{Description} \\
    \midrule
    $\theta,\ \theta'$      & current and proposed weight vectors \\
    $b,\ b'$                & current and proposed bias scalars \\
    $S$                     & scaling constant, $S=10^{x}$ \\
    $\mathcal D_{\mathrm{test}}=\{(x_i,y_i)\}_{i=1}^n$ 
                            & on-chain evaluation dataset \\
    $x_i\in\mathbb R^d,\ y_i\in\{0,1\}$ 
                            & feature vector and label \\
    $f_\theta(\cdot)$       & classifier parameterized by $(\theta,b)$ \\
    $s$                     & stake (in wei) deposited by proposer \\
    $M,M',M_{\mathrm{old}},\widetilde M$ 
                            & metric vectors before/after update \\
    $\Delta t$              & challenge window (e.g., 1-day) \\
    \bottomrule
  \end{tabular}

  \label{Notations}
\end{table}
\noindent\textbf{Proof of Improvement (PoIm).} Let \( f_{\theta} \) denote the deployed classifier with an immutable architecture and on-chain weights \( \theta \) (see Table~\ref{Notations} for notation). The core of our \textit{PoIm} mechanism relies on \( \mathcal{D}_{\text{test}} = \{(x_i, y_i)\}_{i=1}^n \), a canonical evaluation dataset stored directly on-chain. This dataset is intentionally curated to be compact, yet it is representative of critical attack vectors and desired model behaviors rather than being an exhaustive list of all historical transactions. Its manageable size is crucial for enabling efficient on-chain evaluation of proposed model updates within gas limits. Furthermore, to maintain its relevance and resist ossification, \( \mathcal{D}_{\text{test}} \) itself is governed by a decentralized peer-based mechanism (e.g., DAO voting), allowing for agreed-upon additions, modifications, or removals of test samples over time (see Figure \ref{fig:poim_flow}). Let \( \theta' \) represent a submitted update. We define the classifier (i.e., linear) as
$  f_\theta(x) \;=\;
  \mathrm{sign}\!\Bigl(\frac{1}{S}\sum_{i=1}^d \theta_i\,x_i \;+\; b\Bigr)
  \quad\in\{0,1\}.
  \label{eq:classification}$
We define the evaluation function as 
$\mathrm{Eval}(f_\theta, \mathcal{D}_{\text{test}}) \rightarrow (\mathrm{Acc}, \mathrm{F1}, \mathrm{Prec}, \mathrm{Rec})$, 
where all metrics are computed deterministically using Solidity logic and can be challenged.

\noindent\textbf{New Training Submission.}  
Users can propose a new model update by submitting a new training sample directly to the \textit{PoIm} contract, which yields a change (if accepted) in new model weights \( \theta' \) and biases \( b' \), and staking a minimum amount of tokens (or, e.g., any ERC20 token). Each submission must submit a stake, e.g., $s > 0$. The staked ERC20 serves as collateral for the proposal. If accepted, the contributor receives:
$  R \;=\; s \;+\;\sum_{k\in\{\mathrm{Acc},\mathrm{F1},\mathrm{Prec},\mathrm{Rec}\}}
             \alpha_k\,(M'_k - M_k)
  \label{eq:reward}.$
Each coefficient $\alpha_i \geq 0$ reflects the vault's value weighting for each metric improvement. For instance, if the vault has accumulated 1 ETH from failed update attempts, the payout $R$ is proportionally distributed based on the magnitude of improvement across the four metrics.\\
\noindent\textbf{Model Update Acceptance.} 
An update \( \theta' \) is accepted if it improves at least one core metric (accuracy, precision, recall, or F1-score) without degrading any of the other core metrics compared to the current model, based on on-chain evaluation over \( \mathcal{D}_{\text{test}} \). The contract enforces this by ensuring that any accepted model update demonstrates improvement in at least one core metric without degrading others, thereby preventing overfitting a single target (e.g., maximizing recall while degrading precision). This improvement is verified on-chain on $\mathcal{D}_{\text{test}}$, which ensures that the proposed model performs better than the current one.\\
\noindent\textbf{Test Set and Adversarial Update Rollback.}  We integrate an on-chain DAO mechanism that allows stakers (e.g, DeFi platforms) to collectively manage both model rollbacks and $\mathcal{D}_{\text{test}}$ test set updates. In our design, participants stake tokens to gain voting rights to propose and vote on critical actions such as adding or modifying test samples and reverting model weights (to previously committed L1 update) if suspicious behavior is observed (e.g., malicious training sample).

In the case of a challenged update (i.e., one that introduces a loophole but still satisfies the acceptance criteria), a revealed update can be rolled back within a fixed period (e.g., <7 days). If the new model weights worsen in any performance metric compared to the previous metrics, the proposer loses their collateral and the proposal is discarded. 
\section{Evaluation} \label{sec:evaluation} This section details our experimental setup, dataset construction, model configurations, and metrics used to evaluate our proposed framework for on-chain DeFi exploit detection and mitigation across various machine learning architectures. We also present a quantitative comparison against a baseline for on-chain ML.

\begin{figure}[h] 
    \centering
    \includegraphics[width=0.9\linewidth]{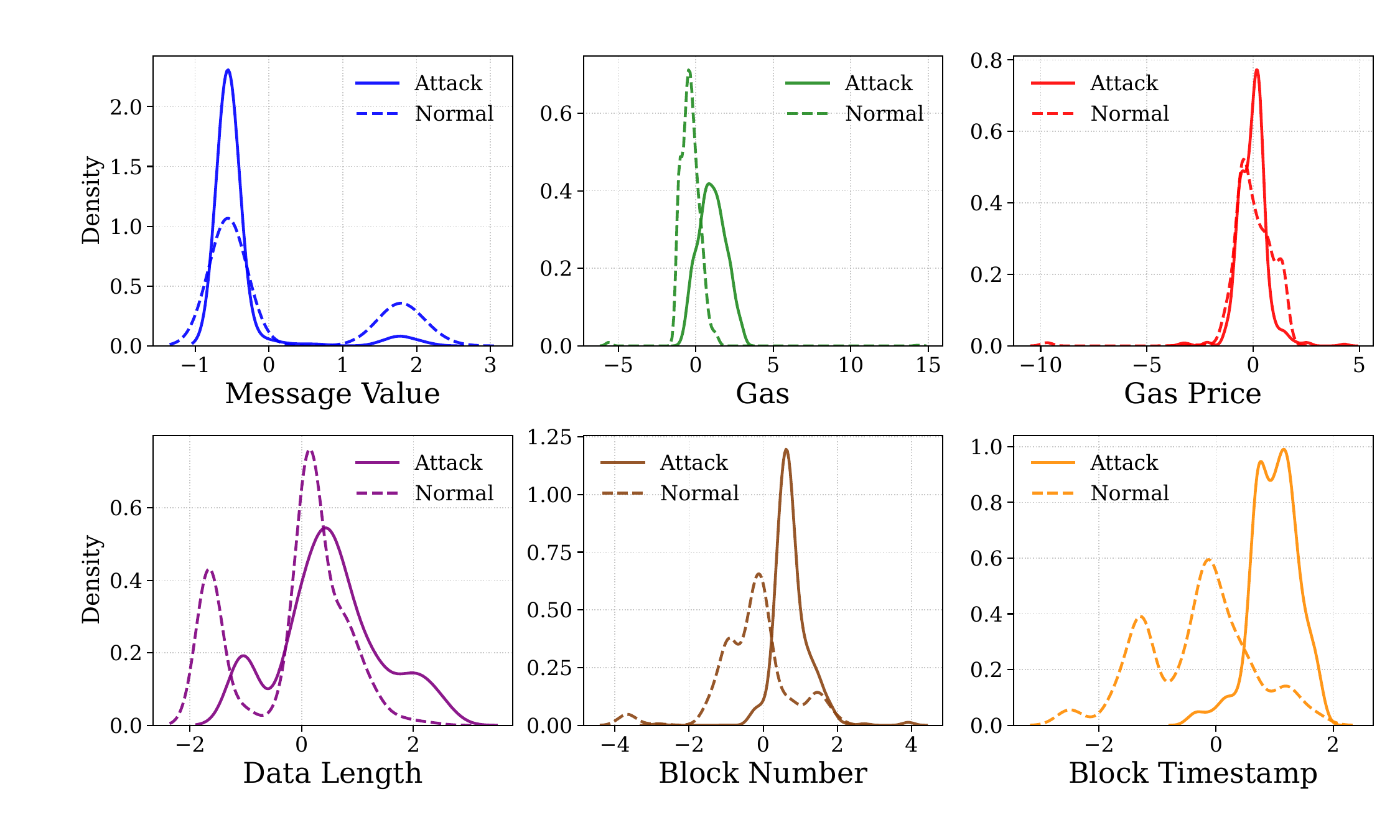} 

    \caption{Distributions of normal and attack transactions.}
    \label{fig:kde} 
\end{figure}

\subsection{Experimental Setup} \label{sec:eval_experimental_setup} All off-chain computations, including model training, parameter generation for Proof-of-Improvement (PoIm) proposals, and performance metric calculations, were developed on a machine equipped with an Intel Core i7 CPU and 16GB of RAM, running Windows 10. Key Python libraries we utilized include \texttt{scikit-learn}~\cite{kramer2016scikit}, \texttt{pandas}~\cite{mckinney2011pandas} (e.g., for data manipulation), \texttt{NumPy}~\cite{harris2020array} (e.g., for numerical operations), \texttt{web3.py} (e.g., for blockchain interaction), \texttt{py-solc-x} (for Solidity compilation), and the \texttt{Z3-solver}~\cite{de2008z3} (e.g., for formal verification logic). 

For blockchain interactions, local L1 and L2 test environments were used. The L2 environment was configured using a Hardhat Network instance forking a live Optimism L2 rollup state (e.g., Optimism Mainnet), facilitating realistic gas calculations and execution behavior mirroring Optimism's characteristics. This L2 environment is primarily used for the decentralized training and \textit{PoIm} mechanism. The L1 environment was simulated using Anvil\cite{foundry}, configured for persistence across experimental runs to reflect a stable mainnet-like chain. \\
\noindent\textbf{Attack Dataset Construction.} A significant challenge in evaluating DeFi exploit detection systems is the general absence of comprehensive, publicly available, raw transaction datasets suitable for behavioral modeling. While some prior work has focused on smart contract code analysis~\cite{zheng2024dappscan, yashavant2022scrawld,zhang2020framework}, readily usable transactional datasets for exploit detection are scarce. To address this, we undertook a meticulous manual collection and multi-stage verification process to construct a robust dataset for this study, covering attacks from 2020 to 2025.
Our initial identification of potential exploits involved a broad survey of diverse sources, including industry security news (e.g., Rekt News \cite{defiRektDatabase}), detailed analyses on technical blogs, discussions on social media platforms (e.g., X), and curated public incident databases such as the DeFi Rekt Database~\cite{defiRektDatabase} and DeFiLlama Exploits Dashboard~\cite{lam}. Incidents and leads gathered from these channels, often further indexed or summarized by resources like DeFiHackLabs~\cite{defihacklabs}, guided our targeted retrieval of the specific transactions that executed each confirmed attack.

These transactions are primarily identified by their hashes on public blockchain explorers (e.g., Etherscan~\cite{etherscan}, Polygonscan \cite{polygonscan}). All transactional records were independently retrieved and subsequently verified through direct blockchain queries via Web3 RPCs \cite{web3js} to ensure authenticity and completeness of call data, receipts, and traces. The primary features extracted for model input include: \texttt{gas} (gas limit provided by sender), \texttt{block.timestamp}, \texttt{func\_selector\_encoded} (a label-encoded representation of the first four bytes of \texttt{msg.data}), \texttt{chain} ID (label-encoded), \texttt{msg.sender} (label-encoded), \texttt{tx.origin} (label-encoded), and \texttt{msg.to} (label-encoded). Numerical features were standardized, and categorical features were label-encoded. Features such as \texttt{block.timestamp} may overfit if the same timestamp is found in both training and testing. To mitigate this risk, we implemented a strict temporal train-test split. Training is exclusively on historical data, and testing is on future (unseen attacks). Therefore, the model generalizes better based on transaction behaviors rather than memorizing temporal artifacts. Transaction traces and contract interactions were analyzed to produce contextual annotations, including \texttt{attack_name}, links to exploited contract source files, incident report URLs, dates, identified root causes, and financial loss in USD (normalized to the time of exploit). This process was applied to transactions across multiple EVM-compatible blockchains, including \textit{Ethereum, Binance Smart Chain (BSC), Polygon, Avalanche, Arbitrum, Fantom, Moonriver, and Base}. Table~\ref{tab:eval_example_dataset} presents an illustrative subset of these attributes.

\vspace{1mm}
\begin{table}[h]
\small
\renewcommand{\arraystretch}{0.88}
\caption{Example of a subset of fields from our exploit transaction dataset. Full transaction records include additional attributes such as detailed EVM context, blockchain metadata, and semantic annotations.}
\label{tab:eval_example_dataset} 
\begin{tabular}{lll}
\toprule
\textbf{Field} & \textbf{Example} & \textbf{Description} \\
\midrule
\texttt{tx\_hash} & 0x78d7...2df4 & Unique identifier of the exploit transaction \\
\texttt{msg.sender} & 0x5aa5...30b7 & Address that initiated the exploit \\
\texttt{msg.value} & 0 & ETH directly transferred in the transaction \\
\texttt{gas\_used} & 298210 & Total gas consumed during execution \\
\texttt{block.timestamp} & 1688462834 & Unix timestamp of the exploit \\
\texttt{root\_cause} & Unchecked external call & Vulnerability exploited in the contract \\
\texttt{loss\_USD} & \$1.94M & Financial loss via unauthorized token transfers \\
\texttt{chain} & Ethereum & Blockchain on which the exploit occurred \\
\bottomrule
\end{tabular}

\noindent\textit{\footnotesize{$^{\ast}$Only a subset of fields is shown here.}}
\end{table}

We observed that an attack can manifest in a single atomic transaction (approximately 71\% of the attacks in our dataset) or across multiple transactions. For multi-transaction attacks, all constituent transactions were grouped and assigned entirely to either the training or testing set to prevent data leakage and maintain the integrity of the attack sequence. The combined dataset of normal and attack transactions was then sorted chronologically by \texttt{block.timestamp}.
Table~\ref{tab:eval_exploit_distribution} summarizes the distribution of the unique exploits that informed the construction of our dataset.
\begin{table}[ht]
\centering
\small
\renewcommand{\arraystretch}{0.88}
\caption{Distribution of \textit{unique} exploits (2020- 2025) and for DeFi financial losses.}
\label{tab:eval_exploit_distribution} 
\begin{tabular}{lrr}
\toprule
\textbf{Category} & \textbf{Count} & \textbf{Loss (USD)} \\
\midrule
Total distinct exploits     & 298 & -- \\
Exploits in training set    & 202 & \$1,877,229,549.86 \\
Exploits in testing set     & 96  & \$1,858,400,900.66 \\
\midrule
\textbf{Total exploits loss} & -- & \textbf{\$3,735,630,450.52} \\
\bottomrule
\end{tabular}
\end{table}

The resulting transaction data capture fine-grained, executable-level behaviors and intricate attacker-victim dynamics. This detailed resolution enables precise detection and classification of exploits and provides a foundation for exploring potential attack discovery. The final curated dataset of 298 unique attack vectors, corresponding to confirmed financial losses exceeding \$3.74 billion (as per Table~\ref{tab:eval_exploit_distribution}), combined with temporally aligned normal transactions, offers a rich and unique foundation for evaluating real-time defense mechanisms and rigorously training on-chain classifiers.

Our dataset construction was guided by the following criteria for a representative sample of DeFi attacks over the past five years (2020 - 2025)\footnote{We collected data up to April 2025.}:\\
\textbf{\ding{172}} \textbf{Economic impact:} Every included incident is a verified on-chain DeFi exploit with financial loss, i.e., $\geq\$10{,}000$. The cumulative loss from the 298 unique exploits considered totals approximately \$3.74 billion. For dataset construction criteria, consistency with the dataset's actual total loss is key. \\
\textbf{\ding{173}} \textbf{Root-cause breadth:} The 298 attacks span at least 12 distinct attack classes (e.g., \emph{business-logic flaws}, \emph{price manipulation oracle attacks}, \emph{arbitrary external calls}, \emph{reentrancy vulnerabilities}, \emph{access-control lapses}), providing comprehensive coverage of common DeFi attack patterns. \\
\textbf{\ding{174}} \textbf{Exploit focus:} Incidents primarily targeting non-fungible tokens (NFTs) were excluded to maintain feature alignment with the fungible-asset liquidity mechanics that are the primary focus of our defense mechanisms. \\
\textbf{\ding{175}} \textbf{Verifiability:} All transactions and affected contracts related to the included exploits are publicly accessible via RPCs on their respective blockchains, enabling deterministic replay, trace analysis, and formal verification of findings.\\
\noindent\textbf{Model Architectures.}
Our framework was evaluated with diverse model architectures, including Multi-Layer Perceptrons (MLPs), Convolutional Neural Networks (CNNs), Recurrent Neural Networks (RNNs), and standard classifiers like Logistic Regression (LogReg), Support Vector Machines (SVM), and Decision Trees (DTs). All models utilized the 7 processed input features detailed previously. These features are observable by smart contracts during execution, enabling real-time attack detection. All models were evaluated with fixed-point parameters (scale $10^{1}$ to $10^{18}$).

\noindent\textbf{Proof-of-Improvement (PoIm) Protocol.}
Each instance starts from baseline parameters and their performance metrics (Accuracy, F1-score, Precision, Recall) on a fixed test set ($\mathcal{D}_{\text{test}}$). An update is accepted only if it improves at least one metric without degrading others. Approved parameters are then eligible for propagation to L1 for inference.

\noindent\textbf{Metrics.}
\label{sec:eval_metrics}The performance of our framework was evaluated across three key dimensions: on-chain inference efficiency, attack detection capabilities, and the model update mechanism via Proof-of-Improvement (PoIm). The specific metrics used within each category are detailed in Table~\ref{tab:eval_evaluation_metrics}.

\vspace{1mm}
\begin{table}[h]
\centering
\small
\caption{Core evaluation metrics for assessing on-chain model performance, detection efficacy, and update mechanisms.}
\label{tab:eval_evaluation_metrics}
\begin{tabular}{lp{9cm}}
\hline
\textbf{Metric} & \textbf{Description} \\
\hline
\multicolumn{2}{c}{\textbf{On-Chain Efficiency}} \\
\hline
L1 Deployment Gas & Total gas to deploy the L1 inference smart contract. \\
L1 Bytecode Size & Size in bytes of the deployed L1 inference contract. \\
L1 Set/Update Params Gas & Gas to set or update model parameters on the L1 contract. \\
L1 Inference Gas & Gas for a single \texttt{classifyOnChain} transaction on L1. \\
L1 Inference Throughput & Number of L1 \texttt{classifyOnChain} transactions processed per second. \\
On-chain/Off-chain Consistency & Proportion of matching predictions between the off-chain (fixed-point) and the on-chain L1 contract for the same inputs and parameters. \\
\hline
\multicolumn{2}{c}{\textbf{Attack Detection Performance}} \\
\hline
Accuracy & Overall fraction of correctly classified (attack or normal) transactions. \\
Precision & Ratio of correctly identified attacks to all transactions flagged as attacks (TP / (TP + FP)). \\
Recall (Sensitivity) & Ratio of correctly identified attacks to all actual attack transactions (TP / (TP + FN)). \\
F1-Score & Harmonic mean of Precision and Recall (2 * (Prec * Rec) / (Prec + Rec)). \\
False Positive Rate (FPR) & Proportion of benign transactions incorrectly classified as attacks (FP / (FP + TN)). \\
\hline
\multicolumn{2}{c}{\textbf{Model Update Mechanism (\textit{PoIm})}} \\
\hline
L2 \textit{PoIm} Update Gas & Total gas for a successful \texttt{proposeUpdate} transaction on the L2 \textit{PoIm} contract. \\

L2-L1 Update Cost & Gas cost to propagate accepted L2 model parameters to the L1 inference contract. \\
\hline
\end{tabular}
\end{table}
\subsection{Performance of On-Chain DeFi Exploit Detection}
\label{sec:eval_detection_performance}
We evaluate the effectiveness of on-chain machine learning classifiers in detecting previously unseen DeFi exploits. Each model was tested on a hold-out test set $\mathcal{D}_{\text{test}}$ comprising real-world attacks and benign transactions, temporally separated from the training set to simulate generalization to zero-day attacks. Core detection metrics are reported in Table~\ref{tab:eval_model_performance_metrics_updated}, and their corresponding financial impact is detailed in Table~\ref{tab:eval_financial_impact_updated}.

We tested a diverse set of models: Logistic Regression (LogReg), Support Vector Machine (SVM), Decision Tree (DT), Multi-Layer Perceptron (MLP), 15 CNN variants (filters $F \in \{2,4,8,10,16\}$, kernel sizes $K \in \{1,4,5\}$), and RNNs with 8 units and with 1 and 7 timesteps (T=1 and T=7). All models utilized a shared 7-feature input vector, preprocessed via standardization and label-encoding as described in Section~\ref{sec:eval_experimental_setup}.

The fully trained CNN variants demonstrated strong detection capabilities, particularly in terms of recall. Many configurations achieved recall $\geq 0.96$, with F1-scores generally ranging from $\sim$0.78 to 0.82, and precision reaching up to 0.8077 (CNN (F4, K1)). For instance, CNN(F4, K4) achieved a high accuracy of 0.9004 and an F1-score of 0.8200, preventing an estimated \$1,857.6M in losses. This performance marks a significant improvement over any preliminary simulations where simpler CNN setups might have exhibited degenerate behavior.

RNN models also showed robust and balanced performance. Specifically, RNN(U8, T1) achieved an accuracy of 0.8517 with a high recall of 0.9792, contributing to \$1,858.2M in prevented losses. The RNN(U8, T7) configuration maintained competitive performance, notably achieving higher precision (0.6607) and a lower false positive count (58 FPs) compared to RNN(U8, T1), with a slight trade-off in recall (0.9479).

Among other classifiers, LogReg and SVM performed well, with SVM achieving the highest AUC in this set (0.9739, see Table~\ref{tab:eval_model_performance_metrics_updated}) and LogReg attaining perfect recall (1.00). The Multi-Layer Perceptron (MLP) also demonstrated strong results, with an accuracy of 0.8665, F1-score of 0.7823, and a high recall of 0.9688. The DecisionTree, while achieving perfect recall, did so at the cost of a significantly higher number of false positives (224 FPs), indicating overfitting to the attack class.
\subsection{Efficiency, Cost, and Consistency}
\label{sec:eval_efficiency_cost}
We analyze the on-chain operational costs, resource utilization, and behavioral consistency of the evaluated machine learning models. All quantitative data discussed refer to Table~\ref{tab:eval_costs_and_poim_summary}, which details L2 and L1 deployment gas, L1 inference gas, contract bytecode sizes, and key Proof-of-Improvement (PoIm) interaction costs where applicable.

\noindent\textbf{Deployment and Inference Gas Costs.}
The on-chain footprint of simpler models like Logistic Regression (LogReg), Support Vector Machines (SVM), Decision Trees (DT), and a Multi-Layer Perceptron (MLP) showcases varying efficiencies. L2 \textit{PoIm} contract deployment gas ranged from approximately 0.88M for the DT to 2.14M for the MLP, while their corresponding L1 inference contracts were lighter. Notably, L1 inference gas for LogReg, SVM, and DT was very low (33k--58k gas). The evaluated MLP required approximately 138k gas for L1 inference, still well within practical limits for on-chain execution.
\begin{figure}[h]
    \centering
    \includegraphics[width=0.7\linewidth]{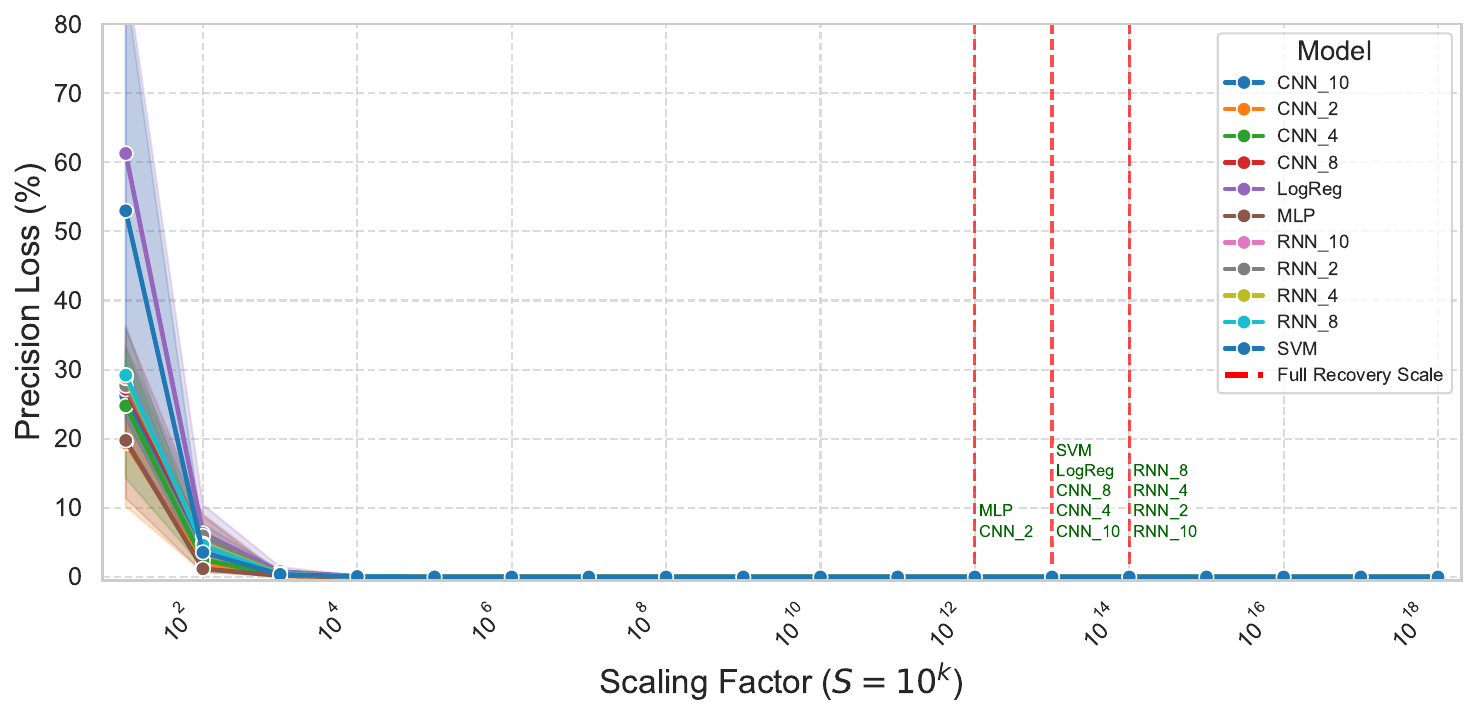}
    \caption{Bitwise consistency for different scaling factors of various models}
    \label{fig:Bitwise}
\end{figure}
Convolutional Neural Network (CNN) variants demonstrated a clear trend where both L2/L1 deployment gas and L1 inference gas scaled with architectural complexity, primarily driven by the number of filters (F) and kernel size (K). For instance, L2 deployment gas ranged from approximately 1.72M for CNN(F2, K1) up to 4.61M for larger configurations like CNN(F16, K1). Similarly, L1 inference gas for these CNNs varied from around 144k (CNN(F2, K1)) to over 969k (CNN(F16, K4)). Bytecode sizes for CNN L1 inference contracts were observed to be consistent for models sharing the same kernel size, as changes in filter count primarily affect the parameter size of the subsequent fully connected layer rather than the convolutional logic structure itself.

For the Recurrent Neural Network (RNN) models evaluated (both with U=8 units), the T=1 configuration (processing all 7 features in a single timestep) incurred higher L2 deployment gas (3.52M) compared to the T=7 configuration (processing 1 feature per 7 timesteps, 2.44M L2 gas). This difference is likely attributable to the larger input-to-hidden weight matrix ($W_{xh}$) in the RNN(U8, T1) model. Conversely, L1 inference gas was substantially higher for RNN(U8, T7) (1.13M gas) than for RNN(U8, T1) (0.56M gas), reflecting the increased number of recurrent steps executed on-chain for the sequence-based input.

The gas cost for transferring updated model parameters from an L2 \textit{PoIm} contract to its L1 inference counterpart varied across model types. For SVM and DT, this L2-L1 update was relatively efficient (96k--231k gas). However, the LogReg and the dynamically trained MLP exhibited higher transfer costs (5.2M--5.7M gas, respectively), likely due to the encoding or size of their complete parameter sets being transferred. For the more complex CNN and RNN models, this L2-L1 parameter transfer for their (typically output layer) updates also showed significant gas consumption, ranging from 2.8M to over 22M gas, underscoring the cost implications of updating larger or more intricate models across layers.
\begin{table}[ht]
\centering
\tiny
\caption{On-Chain costs, sizes, and \textit{PoIm} dynamics for LogReg, SVM, DT, MLP, CNN, and RNN models. \textbf{Ext. Call:} inference by an EVM node.}
\label{tab:eval_costs_and_poim_summary}
\begin{tabular}{@{}l@{\hspace{-2pt}}r@{\hspace{3pt}}r@{\hspace{3pt}}r@{\hspace{2pt}}r@{\hspace{2pt}}r@{\hspace{2pt}}r@{\hspace{2pt}}c@{}}
\toprule
\textbf{Model} & \textbf{Deploy L2} & \textbf{Deploy L1} & \textbf{L1 Inf. Gas} & \textbf{L2 Size} & \textbf{L1 Size} & \textbf{L2-L1 (PoIm)} & \textbf{Ext. Call$^*$} \\
& \textbf{(Gas)} & \textbf{(Gas)} & \textbf{(Gas)} & \textbf{(Bytes)} & \textbf{(Bytes)} & \textbf{(Gas)} & \textbf{(\$)}\\
\midrule
LogReg            & 1,185,010 & 722,020   & 57,603    & 4,597 & 3,044 & 231,168   & 0 \\
SVM           & 1,184,998 & 722,020   & 57,603    & 4,597 & 3,044 & 231,168     & 0 \\
DT            &   878,284 & 435,490   & 33,414    & 3,352 & 1,707 & 96,489      & 0 \\
MLP           & 2,137,045 & 1,116,415 & 138,173   & 7,266 & 4,866 & 5,754,732   & 0 \\
CNN(F2, K1)   & 1,721,688 & 1,536,046 & 143,647   & 6,692 & 5,834 & 2,872,940   & 0 \\
CNN(F4, K1)   & 2,134,766 & 1,949,124 & 244,284   & 6,692 & 5,834 & 4,885,680   & 0 \\
CNN(F8, K1)   & 2,961,086 & 2,775,539 & 445,562   & 6,692 & 5,834 & 8,911,240   & 0 \\
CNN(F10, K1)  & 3,374,357 & 3,188,810 & 546,202   & 6,692 & 5,834 & 10,924,040  & 0 \\
CNN(F16, K1)  & 4,614,143 & 4,428,596 & 848,126   & 6,692 & 5,834 & 16,962,520  & 0 \\
CNN(F2, K4)   & 1,721,688 & 1,536,046 & 158,856   & 6,692 & 5,834 & 3,177,120   & 0 \\
CNN(F4, K4)   & 2,134,778 & 1,949,136 & 274,703   & 6,692 & 5,834 & 5,494,060   & 0 \\
CNN(F8, K4)   & 2,961,098 & 2,775,551 & 506,397   & 6,692 & 5,834 & 10,127,940  & 0 \\
CNN(F10, K4)  & 3,374,369 & 3,188,822 & 622,244   & 6,692 & 5,834 & 12,444,880  & 0 \\
CNN(F16, K4)  & 4,614,167 & 4,428,620 & 969,788   & 6,692 & 5,834 & 19,395,760  & 0 \\
CNN(F2, K5)   & 1,721,688 & 1,536,046 & 150,304   & 6,692 & 5,834 & 3,006,080   & 0 \\
CNN(F4, K5)   & 2,134,778 & 1,949,136 & 257,598   & 6,692 & 5,834 & 5,151,960   & 0 \\
CNN(F8, K5)   & 2,961,086 & 2,775,539 & 472,188   & 6,692 & 5,834 & 9,443,760   & 0 \\
CNN(F10, K5)  & 3,374,345 & 3,188,798 & 579,482   & 6,692 & 5,834 & 11,589,640  & 0 \\
CNN(F16, K5)  & 4,614,143 & 4,428,596 & 901,368   & 6,692 & 5,834 & 18,027,360  & 0 \\
RNN(U8, T1) & 3,520,668 & 3,338,663 & 561,791   & 7,413 & 6,571 & 11,235,820  & 0 \\
RNN(U8, T7) & 2,437,361 & 2,255,356 & 1,131,350 & 7,413 & 6,571 & 22,627,000  & 0 \\
\bottomrule

\end{tabular}
\caption*{\footnotesize\textit{*Read-only RPC calls consume 0 gas on-chain.}}
\end{table}

\vspace{2mm}

\noindent\textbf{Bitwise Consistency Verification.}
We evaluate empirically how fixed-point quantization affects on-chain inference consistency under varying scaling factors from \(10^1\) to \(10^{18}\), using MLP, logistic regression (LogReg), SVM, CNN, and RNN models (Figure \ref{fig:Bitwise}). Precision loss diminishes rapidly as the scale increases, and full recovery (zero bitwise loss across all weights) is achieved at or above \(10^{12}\) for all models. This confirms that fixed-point quantization, when aligned with sufficiently large scaling factors, can reliably preserve inference semantics on-chain across diverse architectures, which was also evaluated in terms of the models' performance.

\begin{figure}[h]
    \centering
    \includegraphics[width=0.95\linewidth]{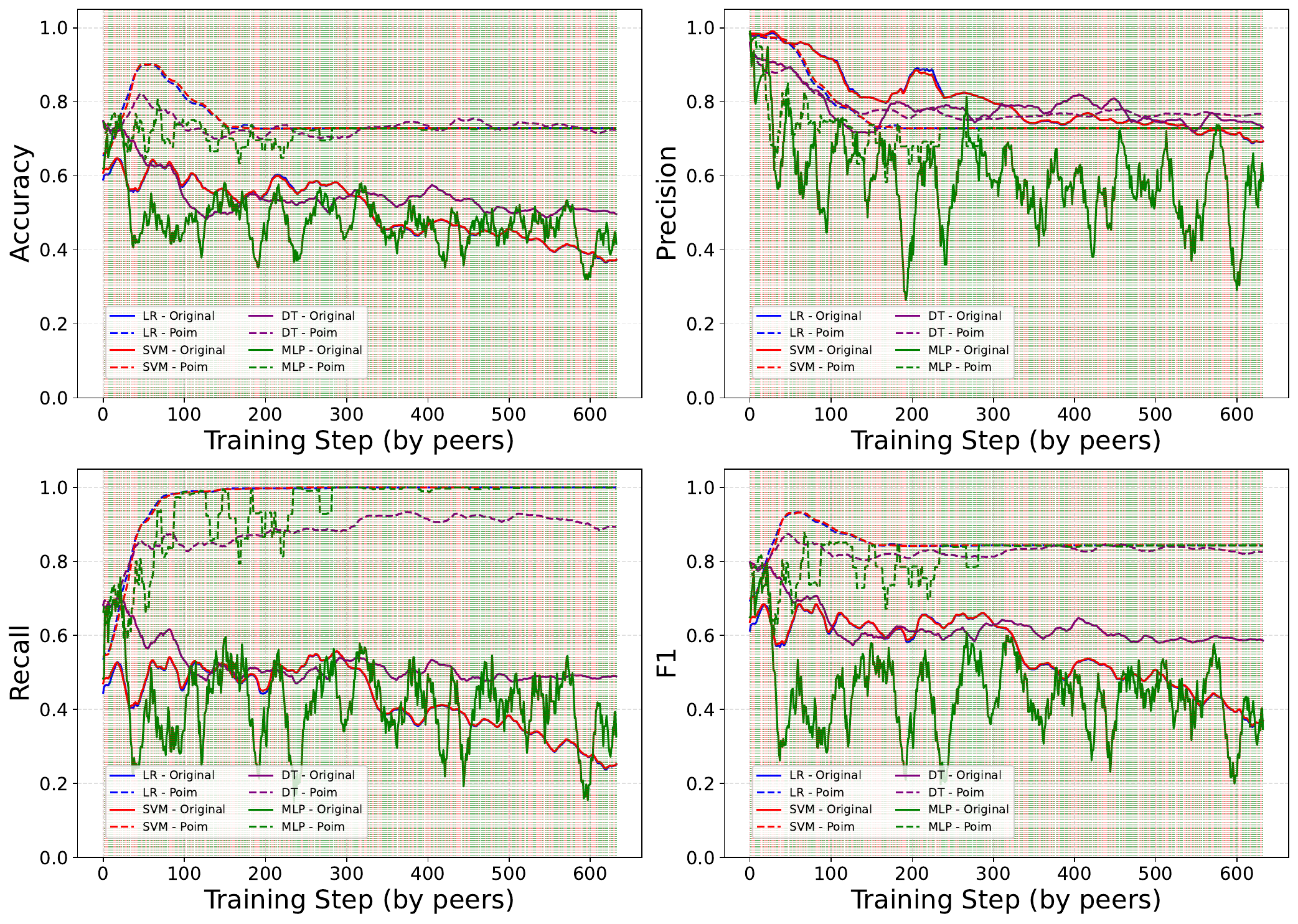}
    \caption{\textit{PoIm} performance under 50\% adversarial training samples. Green vertical lines indicate accepted updates. Red lines indicate rejected ones. \textit{PoIm} stabilizes performance compared to the original training.}
    \label{fig:poim_stress_test}
\end{figure}
\noindent\textbf{\textit{PoIm} Resilience Under Adversarial Training.}
We present \textit{PoIm} resistance against malicious updates via stress testing. We consider bootstrapping the decentralized model with 50 real training samples from both classes, and injecting 50\% of the training data size as malicious or fabricated updates. These malicious updates either flip the class label or inject feature-noise. The 50 real samples are randomly selected from the original attack and normal transaction data to quantify how \textit{PoIm} reacts when training samples degrade one or more metrics, and whether such poisoning affects performance.

\textit{PoIm} learning is incremental. At each new training step, the update is re-evaluated on the full test set. If accepted by \textit{PoIm}, the model is updated; otherwise, the current model is retained. We note that the bootstrapping phase influences how \textit{PoIm} responds to subsequent updates. Overall, \textit{PoIm} maintains more stable performance than the original (unfiltered) training process across all metrics on the same test data, as shown in Figure~\ref{fig:poim_stress_test}. Green vertical bands indicate accepted training samples, while reddish bands denote rejected ones. Notably, original linear models (e.g., logistic regression, SVM) suffer severe degradation without \textit{PoIm} filtering.

\begin{table}[h!]
\centering
\scriptsize
\setlength{\tabcolsep}{4pt}
\caption{Inference throughput per second for (e.g., 10-layer) CNN for different transactions as batches. Ethereum's current TPS $\approx 15$}
\vspace{1pt}
\label{tab:eval_throughput} 
\begin{tabular}{@{}lcc@{}}
\toprule
\textbf{Batch Size} & \textbf{Total Time (s)} & \textbf{Avg Time/Sample (s)} \\
\midrule
1  & 0.0680 & 0.0680 \\
5  & 0.3890 & 0.0778 \\
10 & 0.9150 & 0.0915 \\
20 & 1.7650 & 0.0883 \\
50 & 4.5580 & 0.0912 \\
\bottomrule
\end{tabular}
\end{table}

\noindent\textbf{Inference Throughput.}
To evaluate real-time classification capability for the zero-cost \texttt{eth\_call} tier, we measured inference throughput using our 10-layer CNN. Table \ref{tab:eval_throughput} details these findings, which were obtained by sequentially querying a local Ethereum node. A single transaction is classified in approximately 68ms. While the average per-sample processing time for batches of 5 to 50 transactions stabilizes around 88-91ms, this performance enables a throughput of approximately 11 classifications per second by an off-chain monitoring entity. Such capacity is comparable to Ethereum's current average TPS ($\approx 15$), suitable for timely analysis.

\subsection{Baseline Comparison}
\label{sec:eval_baselines}

We compare our framework against ML2SC~\cite{li2024ml2sc}, which compiles MLP models into smart contracts for on-chain inference. While training in ML2SC is centralized, on-chain ML studies remain sparse. Our evaluation focuses on gas costs and contract behavior. We re-deployed all ML2SC MLP contracts, but observed that their design targets batch processing over a fixed internal dataset of 50 samples. Specifically, their \texttt{classify()} function returns an aggregate result (e.g., count of correct classifications), whereas our framework supports single-instance inference.

Deployment and setup costs differ significantly between our approach and the baseline. Our models embed parameters directly via the constructor and become operational with 1.07M--2.17M gas. In contrast, ML2SC baselines incur higher cumulative setup costs (\textasciitilde{}12.9M--14.0M gas) due to separate contract deployment (1.96M--2.36M gas), parameter setting (0.30M--1.04M gas), and dataset population (\textasciitilde{}10.63M gas for 50 samples), as detailed in Table~\ref{tab:baseline_comparison_summary}. For inference, our framework supports efficient single-input predictions (81k--261k gas) with measurable throughput (\textasciitilde{}7.6--8.0 calls/sec), whereas baseline contracts consume \textasciitilde{}2.0M--2.4M gas per call, reflecting batch evaluation of 50 internal samples. Their \texttt{view}/\texttt{pure} \texttt{classify()} functions are not suited for single-instance transactional inference. Finally, our contracts are smaller in size (\textasciitilde{}4,224 bytes) compared to baseline contracts (\textasciitilde{}9,425--11,151 bytes). The results demonstrate the efficiency and suitability of our framework for single-instance, real-time ML inference on-chain, especially in terms of gas cost and inference flexibility.
\begin{table*}[h!]
\centering
\caption{Our Framework vs. Ml2SC as a baselines~\cite{li2024ml2sc}.}
\label{tab:baseline_comparison_summary}
\resizebox{\textwidth}{!}{
\begin{tabular}{lll}
\toprule
\textbf{Metric} & \textbf{Our Framework} & \textbf{Baseline Method (ML2SC \cite{li2024ml2sc})} \\
\midrule
Deployment Gas (Contract + Params) & 1.07M - 2.17M gas & 1.96M - 2.36M gas (contract code only) \\
Separate Parameter Setting Gas & 0 gas & 0.30M - 1.04M gas \\
Data Setting Gas & N/A & \textbf{\textasciitilde{}10.63M gas} (for 50 internal samples) \\
\textbf{Total Setup Gas} & \textbf{1.07M - 2.17M gas} & \textbf{\textasciitilde{}12.9M - \textasciitilde{}14.0M gas} \\
Bytecode Size (Bytes) & \textasciitilde{}4,224 bytes & \textasciitilde{}9,425 - \textasciitilde{}11,151 bytes \\
Inference Gas (direct classification on chain L1) & \textbf{81k - 261k gas (per single call)} & \textbf{\textasciitilde{}2.0M - \textasciitilde{}2.4M gas (for 50 internal samples)} \\
Output Type & Prediction (0 or 1) & Batch Result (e.g., count like 0 or 14) \\

\bottomrule
\multicolumn{3}{l}{\footnotesize{Baseline \texttt{classify()} output (e.g., 14 (for MLP 1 layer and 1 neuron), 0 for others) indicates the number of correct predictions from its internal batch.}} \\
\end{tabular}
}
\end{table*}

\section{Discussion}
\noindent\textbf{Transaction Class Imbalance Mitigation.} In our decentralized training, the model is updated incrementally by peers through submitted training samples. If normal transactions exceed attack samples by a large margin threshold (e.g., 5x), \textit{PoIm} blocks further normal samples. Only attack samples are accepted until the balance is restored. This prevents skewed updates, which would decrease attack transaction detection accuracy.\\
\noindent\textbf{Trust and Inference Verification.} Our system is designed under the assumption that decentralized participants (i.e., DeFi platforms) perform training on an L2 network. However, we do not assume these participants are honest. To defend against adversarial training data injection, we incorporate per-sample verification and adversarial robustness checks at each training step. Specifically, after each submitted training sample, the \textit{PoIm} evaluates the model against a fixed, immutable (agreed upon by peers) test set using four metrics: accuracy, precision, recall, and F1 score. The update is accepted only if it maintains or improves at least one of these metrics without degrading any others. Otherwise, the training step is discarded, and the data point is excluded from the model. This method is effective in both cases: when malicious training samples are used or when honest user data fails to improve the model’s performance. It ensures that even if malicious actors attempt to inject poisoned or manipulative data, their contributions cannot degrade the classifier's performance. Furthermore, all model training steps are conducted on L2 via a verifiable smart contract, and every accepted update is auditable through its associated state transition and event log. This guarantees transparency and accountability for each model state change. Once a model has reached a finalized state on L2 (e.g., by consensus or performance threshold), its quantized weights are serialized and committed via a cryptographic hash. This hash is propagated to L1 along with the raw weights. The L1 inference contract is a static model (non-trainable) that accepts the weights only if the hash matches, ensuring the integrity of the L2-originated model and protecting the L1 blockchain from tampering or weight substitution attacks. Thus, our architecture enforces correctness and robustness both during training (via metric-based rejection of adversarial updates) and during propagation (via hash-based integrity verification), with no reliance on external oracles, centralized validators.\\
\noindent\textbf{Inference Tiers.} Our two-tier inference architecture imposes a trade-off for protocol designers. The zero-cost tier uses \texttt{eth_call} execution. It can be used for protocols that do not need on-chain inference verifiability, yet get the on-chain verified data, such as wallet interfaces, warning users of potentially malicious transactions without incurring gas fees. For fully verified on-chain defense, protocols integrate the fully on-chain tier, which acts as a gatekeeper (e.g., IPS) by embedding the classification logic interface directly within a state-modifying transaction.\\
\noindent\textbf{Future Work.} The \textit{PoIm} protocol only accepts updates that yield metric improvements. This approach might lead to convergence to a local optimum, where no single micro-update can further improve the model, even if a better global solution exists. Future work could explore other acceptance criteria, such as stochastic policies like simulated annealing \cite{guilmeau2021simulated}, which would permit occasional slightly degrading steps to encourage broader exploration of the model space. Also, \textit{PoIm}, similar to other stake-based governance systems, could be susceptible to centralization if voting rights, represented by tokens (e.g., linear voting), are openly tradable. A malicious actor could accumulate enough stake to influence the update of the test set or roll back stable updates. Mitigating this, for example, through quadratic voting or identity-based mechanisms such as know-your-customer (KYC)~\cite{fabrega2025votingbloc, feichtinger2024sok}, is a potential avenue for future work. Our framework assumes a decentralized and engaged set of participants (e.g., DeFi protocols themselves) whose long-term incentive is to ensure model integrity to protect individual DeFi protocols' funds. Finally, future work may study sophisticated adversarial strategies. This includes addressing game-theoretic risks such as front-running and detecting if latent backdoors in model updates exist.

\color{black}

\section{Related Work}
On-chain AI research seeks transparent, tamper-proof inference but faces the EVM’s fixed-point arithmetic and gas limits~\cite{geren2025blockchain, mafrur2025ai, sham2025generation, li2024ml2sc, ethereum}. Translators from ML to smart contract code such as ML2SC compile MLPs from PyTorch to Solidity, proving feasibility for small models yet incurring high gas per call on complex networks~\cite{sham2025generation, li2024ml2sc}.

For DeFi exploit detection, LookAhead~\cite{ren2024lookahead}, STING~\cite{zhang2023your}, and FlashGuard~\cite{alhaidari2024flashguard} inspect mempools or historical transactions to flag and mitigate attacks. Off-chain placement of these systems introduces latency~\cite{zhang2023your}, misses private-relay transactions, and introduces centralized control. Off-chain ML with on-chain verification is another direction of research. zkML~\cite{chen2024zkml}, for example, attaches zk-SNARK~\cite{petkus2019and} proofs to each inference, preserving privacy but multiplying compute and memory requirements by orders of magnitude~\cite{peng2025survey, ye2025yoimiya, ganescu2024trust}.

Our cryptographic verification is distinct from zkML systems. ZKML frameworks use zero-knowledge proofs (e.g., zk-SNARKs) to verify that a specific computation, such as an ML inference, was executed honestly with a private model~\cite{chen2024zkml}. This provides computational integrity but has substantial overhead. For instance, ZKML needs powerful hardware (up to 1TB of RAM for a distilled GPT-2 model) and can have proving times of nearly an hour~\cite{chen2024zkml}. In contrast, our framework employs a much simpler and more gas-efficient commit-verify scheme. We use a cryptographic hash (keccak256) to ensure the data integrity of the model parameters as they are propagated from L2 to L1. This process guarantees that the model used for inference on L1 is bit-for-bit identical to the one approved by the \textit{PoIm} governance protocol on L2, rather than proving the correctness of each inference itself. Our approach prioritizes provenance and data integrity over computational privacy since the DeFi attacks are public. This makes our approach practical for low-cost, real-time use and suitable for mitigating DeFi attacks.

\color{black}

opML~\cite{conway2024opml} (fraud-proof) treats results as valid unless a verifier proves otherwise, reducing prover cost at the price of economic guarantees. However, it does not provide cryptographic security~\cite{peng2025survey, das2018yoda}. Agatha applies similar fraud proofs to DNNs on Ethereum~\cite{zheng2021agatha}.

Proposals for decentralized model marketplaces, federated learning with ZK privacy, and DAO-based model governance~\cite{alsagheer2023decentralized} either offload heavy compute or evaluate on limited node sets~\cite{conway2024opml}. In contrast, our system shows end-to-end, fully audited inference while remaining within L2 and L1 mainnet gas limits.  Our L1 contracts provide a variety of ML and neural network-based models, bit-exact to the off-chain specification ~\cite{z3prover2025z3, de2008z3} and run efficiently, for example, it incurs only $\approx 57\text{k}$ gas for simple models. More complex non-linear models remain cost-effective, such as MLPs at $\approx 138\text{k}$. L2 \textit{PoIm} governs weight updates via on-chain benchmark tests instead of high-cost zk- or fraud-proof mechanisms. The design keeps the training process and inference decentralized with computation separation (L2 for computation and governance, and L1 purely for inference). Since it is built on smart contracts, it inherently detects transactions coming from any source, such as private relays that seek attack evasion, while allowing continuous community-driven improvement.

\section{Conclusion}
We presented a fully decentralized and verifiable, on-chain ML/DL framework for real-time DeFi exploit detection. Our approach enables classification of transactions at execution time using a deterministic, gas-free inference mechanism embedded in smart contracts. We proposed Proof-of-Improvement (PoIm), a decentralized, stake-based model update protocol that accepts only provably superior updates. The system guarantees inference consistency, bounded gas usage, and resistance to adversarial submissions. Empirical evaluation on 298 real-world DeFi exploits indicates high detection performance and practical feasibility. This work establishes a new model for integrating ML-driven defenses into DeFi protocols with minimal latency, overhead, and maximal decentralization.

\bibliographystyle{plainurl}
\bibliography{bib}

\appendix

\section{L1 Inference Cost on Different Blockchains.}
To understand the practical financial implications of our on-chain inference, we estimated the USD cost of a single L1 inference transaction for a diverse set of models we evaluated across eight blockchains: \textit{Ethereum, Polygon PoS, BNB Smart Chain (BSC), Avalanche C-Chain, Arbitrum One, Optimism, Fantom Opera, Moonriver, and Base}. Table~\ref{tab:appendix_l1_inference_costs_all_models} presents these detailed estimated costs. These calculations utilize the L1 inference gas units from Table~\ref{tab:eval_costs_and_poim_summary} and specific gas prices and native token USD values in the table's caption.

\begin{center}
\tiny 
\renewcommand{\arraystretch}{0.3} 
\begin{longtable}{@{}llcc@{}}
\caption{L1 Inference costs (USD) recomputed using the provided gas prices and token USD values.} \\
\label{tab:appendix_l1_inference_costs_all_models} \\
\toprule
\textbf{Model} & \textbf{Chain} & \textbf{L1 Inf. Gas} & \textbf{Inf. Cost (USD)} \\
\midrule
\endfirsthead
\multicolumn{4}{c}
{{\tablename\ \thetable{} -- continued from previous page}} \\
\toprule
\textbf{Model} & \textbf{Chain} & \textbf{L1 Inf. Gas} & \textbf{Inf. Cost (USD)} \\
\midrule
\endhead
\midrule
\multicolumn{4}{r}{\textit{Continued on next page}} \\
\midrule
\endfoot
\bottomrule
\\

\multicolumn{4}{@{}p{\linewidth}}
{\tiny \textit{As of May 2025, ETH L1: 1.20 Gwei, \$2516.95/ETH. Polygon PoS: 32.0 Gwei, \$0.22/MATIC. BSC: 3.0 Gwei, \$660/BNB. Avalanche: 1.01 Gwei, \$22.50/AVAX. Arbitrum One: 0.0140 Gwei (ETH for gas). Optimism: 0.0010 Gwei (ETH for gas). Moonriver: 3.3 Gwei, \$6.90/MOVR. Fantom Opera: 1.23 Gwei, \$0.58/FTM. Base: 0.0048 Gwei (ETH for gas). ETH price of \$2516.95 used for Arbitrum, Optimism, and Base gas cost calculations.}}

\endlastfoot
LogReg & Ethereum & 57,603 & \$0.173981 \\
LogReg & Polygon PoS & 57,603 & \$0.000406 \\
LogReg & BSC & 57,603 & \$0.114054 \\
LogReg & Avalanche & 57,603 & \$0.001309 \\
LogReg & Arbitrum One & 57,603 & \$0.002030 \\
LogReg & Optimism & 57,603 & \$0.000145 \\
LogReg & Moonriver & 57,603 & \$0.001312 \\
LogReg & Fantom Opera & 57,603 & \$0.000041 \\
LogReg & Base & 57,603 & \$0.000696 \\
\midrule
SVM & Ethereum & 57,603 & \$0.173981 \\
SVM & Polygon PoS & 57,603 & \$0.000406 \\
SVM & BSC & 57,603 & \$0.114054 \\
SVM & Avalanche & 57,603 & \$0.001309 \\
SVM & Arbitrum One & 57,603 & \$0.002030 \\
SVM & Optimism & 57,603 & \$0.000145 \\
SVM & Moonriver & 57,603 & \$0.001312 \\
SVM & Fantom Opera & 57,603 & \$0.000041 \\
SVM & Base & 57,603 & \$0.000696 \\
\midrule
DT & Ethereum & 33,414 & \$0.100922 \\
DT & Polygon PoS & 33,414 & \$0.000235 \\
DT & BSC & 33,414 & \$0.066160 \\
DT & Avalanche & 33,414 & \$0.000759 \\
DT & Arbitrum One & 33,414 & \$0.001177 \\
DT & Optimism & 33,414 & \$0.000084 \\
DT & Moonriver & 33,414 & \$0.000761 \\
DT & Fantom Opera & 33,414 & \$0.000024 \\
DT & Base & 33,414 & \$0.000404 \\
\midrule
MLP & Ethereum & 138,173 & \$0.417329 \\
MLP & Polygon PoS & 138,173 & \$0.000973 \\
MLP & BSC & 138,173 & \$0.273583 \\
MLP & Avalanche & 138,173 & \$0.003140 \\
MLP & Arbitrum One & 138,173 & \$0.004869 \\
MLP & Optimism & 138,173 & \$0.000348 \\
MLP & Moonriver & 138,173 & \$0.003146 \\
MLP & Fantom Opera & 138,173 & \$0.000099 \\
MLP & Base & 138,173 & \$0.001669 \\
\midrule
CNN(F2, K1) & Ethereum & 143,647 & \$0.433863 \\
CNN(F2, K1) & Polygon PoS & 143,647 & \$0.001011 \\
CNN(F2, K1) & BSC & 143,647 & \$0.284421 \\
CNN(F2, K1) & Avalanche & 143,647 & \$0.003264 \\
CNN(F2, K1) & Arbitrum One & 143,647 & \$0.005062 \\
CNN(F2, K1) & Optimism & 143,647 & \$0.000362 \\
CNN(F2, K1) & Moonriver & 143,647 & \$0.003271 \\
CNN(F2, K1) & Fantom Opera & 143,647 & \$0.000102 \\
CNN(F2, K1) & Base & 143,647 & \$0.001735 \\
\midrule
CNN(F4, K1) & Ethereum & 244,284 & \$0.737821 \\
CNN(F4, K1) & Polygon PoS & 244,284 & \$0.001720 \\
CNN(F4, K1) & BSC & 244,284 & \$0.483682 \\
CNN(F4, K1) & Avalanche & 244,284 & \$0.005551 \\
CNN(F4, K1) & Arbitrum One & 244,284 & \$0.008608 \\
CNN(F4, K1) & Optimism & 244,284 & \$0.000615 \\
CNN(F4, K1) & Moonriver & 244,284 & \$0.005562 \\
CNN(F4, K1) & Fantom Opera & 244,284 & \$0.000174 \\
CNN(F4, K1) & Base & 244,284 & \$0.002951 \\
\midrule
CNN(F8, K1) & Ethereum & 445,562 & \$1.345749 \\
CNN(F8, K1) & Polygon PoS & 445,562 & \$0.003137 \\
CNN(F8, K1) & BSC & 445,562 & \$0.882213 \\
CNN(F8, K1) & Avalanche & 445,562 & \$0.010125 \\
CNN(F8, K1) & Arbitrum One & 445,562 & \$0.015700 \\
CNN(F8, K1) & Optimism & 445,562 & \$0.001121 \\
CNN(F8, K1) & Moonriver & 445,562 & \$0.010145 \\
CNN(F8, K1) & Fantom Opera & 445,562 & \$0.000318 \\
CNN(F8, K1) & Base & 445,562 & \$0.005383 \\
\midrule
CNN(F10, K1) & Ethereum & 546,202 & \$1.649716 \\
CNN(F10, K1) & Polygon PoS & 546,202 & \$0.003845 \\
CNN(F10, K1) & BSC & 546,202 & \$1.081480 \\
CNN(F10, K1) & Avalanche & 546,202 & \$0.012412 \\
CNN(F10, K1) & Arbitrum One & 546,202 & \$0.019247 \\
CNN(F10, K1) & Optimism & 546,202 & \$0.001375 \\
CNN(F10, K1) & Moonriver & 546,202 & \$0.012437 \\
CNN(F10, K1) & Fantom Opera & 546,202 & \$0.000390 \\
CNN(F10, K1) & Base & 546,202 & \$0.006599 \\
\midrule
CNN(F16, K1) & Ethereum & 848,126 & \$2.561629 \\
CNN(F16, K1) & Polygon PoS & 848,126 & \$0.005971 \\
CNN(F16, K1) & BSC & 848,126 & \$1.679289 \\
CNN(F16, K1) & Avalanche & 848,126 & \$0.019274 \\
CNN(F16, K1) & Arbitrum One & 848,126 & \$0.029886 \\
CNN(F16, K1) & Optimism & 848,126 & \$0.002135 \\
CNN(F16, K1) & Moonriver & 848,126 & \$0.019312 \\
CNN(F16, K1) & Fantom Opera & 848,126 & \$0.000605 \\
CNN(F16, K1) & Base & 848,126 & \$0.010247 \\
\midrule
CNN(F2, K4) & Ethereum & 158,856 & \$0.479799 \\
CNN(F2, K4) & Polygon PoS & 158,856 & \$0.001118 \\
CNN(F2, K4) & BSC & 158,856 & \$0.314535 \\
CNN(F2, K4) & Avalanche & 158,856 & \$0.003610 \\
CNN(F2, K4) & Arbitrum One & 158,856 & \$0.005598 \\
CNN(F2, K4) & Optimism & 158,856 & \$0.000400 \\
CNN(F2, K4) & Moonriver & 158,856 & \$0.003617 \\
CNN(F2, K4) & Fantom Opera & 158,856 & \$0.000113 \\
CNN(F2, K4) & Base & 158,856 & \$0.001919 \\
\midrule
CNN(F4, K4) & Ethereum & 274,703 & \$0.829696 \\
CNN(F4, K4) & Polygon PoS & 274,703 & \$0.001934 \\
CNN(F4, K4) & BSC & 274,703 & \$0.543912 \\
CNN(F4, K4) & Avalanche & 274,703 & \$0.006243 \\
CNN(F4, K4) & Arbitrum One & 274,703 & \$0.009680 \\
CNN(F4, K4) & Optimism & 274,703 & \$0.000691 \\
CNN(F4, K4) & Moonriver & 274,703 & \$0.006255 \\
CNN(F4, K4) & Fantom Opera & 274,703 & \$0.000196 \\
CNN(F4, K4) & Base & 274,703 & \$0.003319 \\
\midrule
CNN(F8, K4) & Ethereum & 506,397 & \$1.529491 \\
CNN(F8, K4) & Polygon PoS & 506,397 & \$0.003565 \\
CNN(F8, K4) & BSC & 506,397 & \$1.002666 \\
CNN(F8, K4) & Avalanche & 506,397 & \$0.011508 \\
CNN(F8, K4) & Arbitrum One & 506,397 & \$0.017844 \\
CNN(F8, K4) & Optimism & 506,397 & \$0.001275 \\
CNN(F8, K4) & Moonriver & 506,397 & \$0.011531 \\
CNN(F8, K4) & Fantom Opera & 506,397 & \$0.000361 \\
CNN(F8, K4) & Base & 506,397 & \$0.006118 \\
\midrule
CNN(F10, K4) & Ethereum & 622,244 & \$1.879388 \\
CNN(F10, K4) & Polygon PoS & 622,244 & \$0.004381 \\
CNN(F10, K4) & BSC & 622,244 & \$1.232043 \\
CNN(F10, K4) & Avalanche & 622,244 & \$0.014140 \\
CNN(F10, K4) & Arbitrum One & 622,244 & \$0.021926 \\
CNN(F10, K4) & Optimism & 622,244 & \$0.001566 \\
CNN(F10, K4) & Moonriver & 622,244 & \$0.014168 \\
CNN(F10, K4) & Fantom Opera & 622,244 & \$0.000444 \\
CNN(F10, K4) & Base & 622,244 & \$0.007518 \\
\midrule
CNN(F16, K4) & Ethereum & 969,788 & \$2.929089 \\
CNN(F16, K4) & Polygon PoS & 969,788 & \$0.006827 \\
CNN(F16, K4) & BSC & 969,788 & \$1.920180 \\
CNN(F16, K4) & Avalanche & 969,788 & \$0.022038 \\
CNN(F16, K4) & Arbitrum One & 969,788 & \$0.034173 \\
CNN(F16, K4) & Optimism & 969,788 & \$0.002441 \\
CNN(F16, K4) & Moonriver & 969,788 & \$0.022082 \\
CNN(F16, K4) & Fantom Opera & 969,788 & \$0.000692 \\
CNN(F16, K4) & Base & 969,788 & \$0.011716 \\
\midrule
CNN(F2, K5) & Ethereum & 150,304 & \$0.453969 \\
CNN(F2, K5) & Polygon PoS & 150,304 & \$0.001058 \\
CNN(F2, K5) & BSC & 150,304 & \$0.297602 \\
CNN(F2, K5) & Avalanche & 150,304 & \$0.003416 \\
CNN(F2, K5) & Arbitrum One & 150,304 & \$0.005296 \\
CNN(F2, K5) & Optimism & 150,304 & \$0.000378 \\
CNN(F2, K5) & Moonriver & 150,304 & \$0.003422 \\
CNN(F2, K5) & Fantom Opera & 150,304 & \$0.000107 \\
CNN(F2, K5) & Base & 150,304 & \$0.001816 \\
\midrule
CNN(F4, K5) & Ethereum & 257,598 & \$0.778034 \\
CNN(F4, K5) & Polygon PoS & 257,598 & \$0.001813 \\
CNN(F4, K5) & BSC & 257,598 & \$0.510044 \\
CNN(F4, K5) & Avalanche & 257,598 & \$0.005854 \\
CNN(F4, K5) & Arbitrum One & 257,598 & \$0.009077 \\
CNN(F4, K5) & Optimism & 257,598 & \$0.000648 \\
CNN(F4, K5) & Moonriver & 257,598 & \$0.005866 \\
CNN(F4, K5) & Fantom Opera & 257,598 & \$0.000184 \\
CNN(F4, K5) & Base & 257,598 & \$0.003112 \\
\midrule
CNN(F8, K5) & Ethereum & 472,188 & \$1.426168 \\
CNN(F8, K5) & Polygon PoS & 472,188 & \$0.003324 \\
CNN(F8, K5) & BSC & 472,188 & \$0.934932 \\
CNN(F8, K5) & Avalanche & 472,188 & \$0.010730 \\
CNN(F8, K5) & Arbitrum One & 472,188 & \$0.016639 \\
CNN(F8, K5) & Optimism & 472,188 & \$0.001188 \\
CNN(F8, K5) & Moonriver & 472,188 & \$0.010752 \\
CNN(F8, K5) & Fantom Opera & 472,188 & \$0.000337 \\
CNN(F8, K5) & Base & 472,188 & \$0.005705 \\
\midrule
CNN(F10, K5) & Ethereum & 579,482 & \$1.750233 \\
CNN(F10, K5) & Polygon PoS & 579,482 & \$0.004080 \\
CNN(F10, K5) & BSC & 579,482 & \$1.147374 \\
CNN(F10, K5) & Avalanche & 579,482 & \$0.013169 \\
CNN(F10, K5) & Arbitrum One & 579,482 & \$0.020419 \\
CNN(F10, K5) & Optimism & 579,482 & \$0.001459 \\
CNN(F10, K5) & Moonriver & 579,482 & \$0.013195 \\
CNN(F10, K5) & Fantom Opera & 579,482 & \$0.000413 \\
CNN(F10, K5) & Base & 579,482 & \$0.007001 \\
\midrule
CNN(F16, K5) & Ethereum & 901,368 & \$2.722438 \\
CNN(F16, K5) & Polygon PoS & 901,368 & \$0.006346 \\
CNN(F16, K5) & BSC & 901,368 & \$1.784709 \\
CNN(F16, K5) & Avalanche & 901,368 & \$0.020484 \\
CNN(F16, K5) & Arbitrum One & 901,368 & \$0.031762 \\
CNN(F16, K5) & Optimism & 901,368 & \$0.002269 \\
CNN(F16, K5) & Moonriver & 901,368 & \$0.020524 \\
CNN(F16, K5) & Fantom Opera & 901,368 & \$0.000643 \\
CNN(F16, K5) & Base & 901,368 & \$0.010890 \\
\midrule
RNN(U8, T1) & Ethereum & 561,791 & \$1.696800 \\
RNN(U8, T1) & Polygon PoS & 561,791 & \$0.003955 \\
RNN(U8, T1) & BSC & 561,791 & \$1.112346 \\
RNN(U8, T1) & Avalanche & 561,791 & \$0.012767 \\
RNN(U8, T1) & Arbitrum One & 561,791 & \$0.019796 \\
RNN(U8, T1) & Optimism & 561,791 & \$0.001414 \\
RNN(U8, T1) & Moonriver & 561,791 & \$0.012792 \\
RNN(U8, T1) & Fantom Opera & 561,791 & \$0.000401 \\
RNN(U8, T1) & Base & 561,791 & \$0.006787 \\
\midrule
RNN(U8, T7) & Ethereum & 1,131,350 & \$3.417062 \\
RNN(U8, T7) & Polygon PoS & 1,131,350 & \$0.007965 \\
RNN(U8, T7) & BSC & 1,131,350 & \$2.240073 \\
RNN(U8, T7) & Avalanche & 1,131,350 & \$0.025710 \\
RNN(U8, T7) & Arbitrum One & 1,131,350 & \$0.039866 \\
RNN(U8, T7) & Optimism & 1,131,350 & \$0.002848 \\
RNN(U8, T7) & Moonriver & 1,131,350 & \$0.025761 \\
RNN(U8, T7) & Fantom Opera & 1,131,350 & \$0.000807 \\
RNN(U8, T7) & Base & 1,131,350 & \$0.013668 \\
\end{longtable}
\end{center}

\section{Detection and Financial Impact}
First, Table \ref{tab:eval_model_performance_metrics_updated} presents the core detection metrics, including accuracy, F1-score, precision, recall, AUC, and False Positive Rate (FPR), for each model when evaluated against previously unseen DeFi exploits from our curated dataset. These metrics quantify the effectiveness of the classifiers in distinguishing between malicious and benign transactions. The recall is computed based on the 96 unique attacks (120 attack transactions) in the test set, while the FPR is computed over the 376 normal transactions in the same set.

\begin{table}[ht]
\centering
\tiny
\caption{Model performance metrics for detection of unseen attacks. Recall is computed as TP\% of 96 unique attacks (120 attack transactions). FP\% is computed over the 376 normal transactions. The test set has 472 transactions in total.} 
\label{tab:eval_model_performance_metrics_updated}
\begin{tabular}{@{}lrrrrrr@{}} 
\toprule
\textbf{Model} & \textbf{Acc.} & \textbf{F1} & \textbf{Prec.} & \textbf{Recall (TP\%)} & \textbf{AUC} & \textbf{FP\%} \\
\midrule
LogReg         & 0.8538 & 0.7738 & 0.6310 & 1.0000 & 0.9516 & 0.1862 \\
DecisionTree   & 0.5381 & 0.5169 & 0.3485 & 1.0000 & 0.6836 & 0.5957 \\
CNN(F10, K4)   & 0.8750 & 0.7950 & 0.6705 & 0.9792 & 0.9552 & 0.1543 \\
CNN(F10, K5)   & 0.8729 & 0.7907 & 0.6686 & 0.9792 & 0.9611 & 0.1569 \\
CNN(F16, K1)   & 0.8665 & 0.7840 & 0.6554 & 0.9792 & 0.9665 & 0.1649 \\
CNN(F8, K4)    & 0.8665 & 0.7840 & 0.6554 & 0.9792 & 0.9505 & 0.1649 \\
CNN(F16, K5)   & 0.8644 & 0.7811 & 0.6517 & 0.9792 & 0.9576 & 0.1676 \\
RNN(U8, T1)       & 0.8517 & 0.7656 & 0.6304 & 0.9792 & 0.9391 & 0.1835 \\
CNN(F8, K5)    & 0.8729 & 0.7893 & 0.6706 & 0.9688 & 0.9495 & 0.1516 \\
CNN(F16, K4)   & 0.8686 & 0.7836 & 0.6628 & 0.9688 & 0.9679 & 0.1569 \\
MLP            & 0.8665 & 0.7823 & 0.6571 & 0.9688 & 0.9666 & 0.1622 \\
CNN(F10, K1)   & 0.8665 & 0.7823 & 0.6571 & 0.9688 & 0.9658 & 0.1622 \\
SVM            & 0.8814 & 0.8000 & 0.6867 & 0.9583 & 0.9739 & 0.1383 \\
CNN(F8, K1)    & 0.8792 & 0.7960 & 0.6848 & 0.9583 & 0.9554 & 0.1410 \\
RNN(U8, T7)       & 0.8644 & 0.7736 & 0.6607 & 0.9479 & 0.9728 & 0.1543 \\
CNN(F4, K4)    & 0.9004 & 0.8200 & 0.7415 & 0.9271 & 0.9577 & 0.1037 \\
CNN(F4, K1)    & 0.8877 & 0.7962 & 0.7338 & 0.8750 & 0.9450 & 0.1011 \\
CNN(F2, K1)    & 0.8072 & 0.6789 & 0.5780 & 0.8125 & 0.8431 & 0.1915 \\
CNN(F2, K4)    & 0.7627 & 0.3055 & 0.5750 & 0.2083 & 0.8097 & 0.0479 \\
CNN(F4, K5)    & 0.7775 & 0.2205 & 0.9333 & 0.1250 & 0.8626 & 0.0027 \\
\bottomrule
\end{tabular}
\end{table}

We also present the financial savings resulting from implementing each on-chain model (see Table \ref{tab:eval_financial_impact_updated}).

\begin{table}[ht]
\centering
\footnotesize 
\caption{Financial Impact of Detection (Ordered by Loss Prevented). \textbf{F:} \#filters and \textbf{K:} \#kernels }
\label{tab:eval_financial_impact_updated}
\begin{tabular}{@{}lrr@{}}
\toprule
\textbf{Model} & \textbf{Loss Prevented (\$)} & \textbf{Loss Missed (\$)} \\
\midrule
DecisionTree   &   1,858.4M &       0.0M \\
LogReg         &   1,858.4M &       0.0M \\
CNN(F10, K4)   &   1,858.2M &       0.2M \\
CNN(F10, K5)   &   1,858.2M &       0.2M \\
CNN(F16, K1)   &   1,858.2M &       0.2M \\
CNN(F16, K5)   &   1,858.2M &       0.2M \\
CNN(F8, K4)    &   1,858.2M &       0.2M \\
CNN(F10, K1)   &   1,858.2M &       0.2M \\
RNN(U8, T1)       &   1,858.2M &       0.2M \\
CNN(F16, K4)   &   1,858.2M &       0.2M \\
MLP            &   1,858.2M &       0.2M \\
CNN(F8, K1)    &   1,858.1M &       0.3M \\
SVM            &   1,858.1M &       0.3M \\
CNN(F8, K5)    &   1,857.8M &       0.6M \\
RNN(U8, T7)       &   1,857.7M &       0.7M \\
CNN(F4, K4)    &   1,857.6M &       0.8M \\
CNN(F2, K1)    &   1,805.0M &      53.4M \\
CNN(F4, K1)    &   1,774.5M &      83.9M \\
CNN(F2, K4)    &     145.3M &   1,713.1M \\
CNN(F4, K5)    &     109.5M &   1,748.9M \\
\bottomrule
\end{tabular}
\end{table}

\section{Root Causes of Attacks}\label{RC}
We classify the root causes of these attacks into five empirically supported categories~\cite{defiRektDatabase, defihacklabs,lam}: 
\begin{itemize}
    \item \textbf{Access control failures:} Contracts lack proper access modifiers (e.g., \texttt{onlyOwner}) or expose privileged functions to unauthorized users (e.g., Dexible, GFOX)

    \item \textbf{Business logic flaws:} Valid transactions result in unintended outcomes, such as undercollateralized loans or unchecked withdrawals. (e.g., Euler, Platypus, PineProtocol).

    \item \textbf{Slippage and oracle manipulation:} Exploits that manipulate oracles, liquidity curves, or swap sequences to skew pricing (e.g., KyberSwap)

    \item \textbf{Unchecked external calls and delegatecall misuse:} Insecure call forwarding or forged approvals lead to attacker-controlled execution. (e.g., Seneca, RabbyRouter, RevertFinance).

    \item \textbf{Storage layout collisions:} Overlapping storage slots in upgradeable contracts results in overwrite of admin roles or token balances (e.g., Telcoin, EFVault).
\end{itemize}
Transaction labeling is derived directly from exploit disclosures and does not rely on heuristic classification.

\section{Consistency Verification of On-Chain Models}\label{sec:appendix_pseudocode_final}

This appendix details the pseudocode for an algorithm designed to ensure the consistency of model parameters and inference behavior when deployed and managed on a blockchain. It outlines procedures for initializing model registries on Layer 2 (L2), updating models via a Proof-of-Improvement (\textit{PoIm}) mechanism where the L2 contract evaluates proposals on-chain, transferring model parameters to Layer 1 (L1) for inference, and verifying the consistency of on-chain predictions. The algorithms (1-4) reflect the logic for maintaining data integrity and computational accuracy in a decentralized ML, consistent with the mechanisms described in the main paper, and is applicable to various parametric model architectures.

\begin{longtable}{ >{\raggedright\arraybackslash}p{0.25\textwidth} p{0.7\textwidth} }
\caption{Summary of notations}. 
\footnotesize 
\label{tab:notations_final}\\
\hline
\textbf{Symbol} & \textbf{Description} \\
\hline
\endfirsthead
\caption[]{Summary of Notations (Continued)} \\
\hline
\textbf{Symbol} & \textbf{Description} \\
\hline
\endhead
\hline
\endfoot
\hline
\endlastfoot
$\Moffchain$ & Off-chain Machine Learning model. \\
$\Theta = \{W, b, L\}$ & Model parameters, where $W$ are weights, $b$ are biases, and $L$ are layer sizes (or other structural parameters). \\
$P = \{Acc, F1, Prec, Rec\}$ & Model performance metrics (Accuracy, F1-score, Precision, Recall). \\
$\Sw$ & Scaling factor for converting $\Theta$ components (weights, biases) to fixed-point integers. Also referred to as $S$ in the main paper. \\
$\Sm$ & Scaling factor for converting $P$ components to fixed-point integers. \\
$\Dtest$ & Canonical on-chain evaluation dataset used by $\CPoIm$ \\
$\CPoIm$ & Deployed L2 smart contract for \textit{PoIm}. Stores $\Thetabestfixed$ and $P_{\text{best\_fixed}}$. \\
$\CInfer$ & Deployed L1 Smart Contract for Inference. Stores $\Thetainferfixed$. \\
$\Thetafixed$ & Generic fixed-point representation of model parameters. \\
$P_{\text{fixed}}$ & Generic fixed-point representation of performance metrics (often used for off-chain calculated metrics). \\
$\Pnewonchainfixed$ & Fixed-point performance metrics of a new model, as calculated on-chain by $\CPoIm$ using $\Dtest$. \\
$\Thetabestfixed$ & Current best fixed-point parameters stored in $\CPoIm$. \\
$P_{\text{best\_fixed}}$ & Current best fixed-point metrics stored in $\CPoIm$ (derived from on-chain). \\
$\Thetainferfixed$ & Fixed-point parameters stored in $\CInfer$ for inference. \\
$\Xsample$ & A sample input feature vector. \\
$\Xsamplefixed$ & Fixed-point representation of $\Xsample$. \\
$\Yoffchain$ & Prediction output from the off-chain model. \\
$\Yonchain$ & Prediction output from the on-chain model. \\
$\tofixed{val}{scale}$ & Conceptual function to convert $val$ to fixed-point using $scale$. \\
$\fromfixed{val}{scale}$ & Conceptual function to convert fixed-point $val$ back using $scale$. \\
$\integercast{val}$ & Conceptual function to cast $val$ to an integer type. \\
\end{longtable}

\vspace{-3pt}

\begin{algorithm}
\caption{Part 1: InitializeL2PoImContract$(\Moffchain^{(0)}, \Dtest\_{\text{off-chain}}, \Sw, \Sm, \CPoIm\_{\text{address}})$}
\label{alg:part1_init_l2_final}
\begin{algorithmic}[1]
    \State Extract initial parameters $\Theta^{(0)}_{\text{off-chain}}$ from $\Moffchain^{(0)}$.
    \State Convert parameters to fixed-point: $\Theta^{(0)}_{\text{fixed}} \leftarrow \{\tofixed{W^{(0)}}{\Sw}, \tofixed{b^{(0)}}{\Sw}, L^{(0)}\}$.
    \State $P^{(0)}\_{\text{expected\_off-chain}} \leftarrow \Moffchain^{(0)}.\text{evaluate}(\Dtest\_{\text{off-chain}})$.
    \State Deploy $\CPoIm$ to L2. The constructor takes $(\Theta^{(0)}_{\text{fixed}}, \Sw, \Sm)$ and internally evaluates $\Theta^{(0)}_{\text{fixed}}$ on the on-chain $\Dtest$ to establish the initial $P_{\text{best\_fixed}}$.
    \State Let $P^{(0)}_{\text{on-chain\_eval}} \leftarrow \text{metrics resulting from } \CPoIm \text{'s internal initial evaluation}$.
    \State \textbf{Verification 1 (Initial State Consistency on L2):}
    \State \quad Retrieve current best parameters from $\CPoIm$: $\Thetabestfixed \leftarrow \CPoIm.\text{getCurrentModelParameters}()$.
    \State \quad Retrieve current best metrics from $\CPoIm$: $P_{\text{best\_fixed}} \leftarrow \CPoIm.\text{getCurrentModelMetrics}()$.
    \State \quad Assert that $\Thetabestfixed = \Theta^{(0)}_{\text{fixed}}$.
    \State \quad Assert that $P_{\text{best\_fixed}} = P^{(0)}_{\text{on-chain\_eval}}$.
\end{algorithmic}
\end{algorithm}

\vspace{-2pt}

\begin{algorithm}
\caption{Part 2: ProposeAndUpdateL2PoIm$(\Moffchain^{(\text{new})}, \CPoIm, \Sw)$}
\label{alg:part2_update_l2_final}
\begin{algorithmic}[1]
    \State Extract new parameters $\Theta^{(\text{new})}_{\text{off-chain}}$ from $\Moffchain^{(\text{new})}$.
    \State Convert new parameters to fixed-point: $\Theta^{(\text{new})}_{\text{fixed}} \leftarrow \{\tofixed{W^{(\text{new})}}{\Sw}, \tofixed{b^{(\text{new})}}{\Sw}, L^{(\text{new})}\}$.
    \State Store the current on-chain best parameters and metrics (before transaction):
    \State $\Thetabestfixedold \leftarrow \CPoIm.\text{getCurrentModelParameters}()$.
    \State $P_{\text{best\_fixed\_old}} \leftarrow \CPoIm.\text{getCurrentModelMetrics}()$.
    \State $\text{tx\_receipt} \leftarrow \CPoIm.\text{proposeModelUpdate}(\Theta^{(\text{new})}_{\text{fixed}}, \text{stake } s)$.
    \State $\text{accepted} \leftarrow \text{determine\_from\_event\_or\_return}(\text{tx\_receipt})$.
    \State Let $\Pnewonchainfixed\_{\text{from\_event\_if\_accepted}}$ be the new metrics if update was accepted.
    \If{\text{accepted}}
        \State \textbf{Verification 2.1 (Accepted Update - State Change):}
        \State \quad Retrieve current parameters from $\CPoIm$: $\Thetabestfixedcurrent$.
        \State \quad Retrieve current metrics from $\CPoIm$: $P_{\text{best\_fixed\_current}}$.
        \State \quad Assert that $\Thetabestfixedcurrent = \Theta^{(\text{new})}_{\text{fixed}}$.
        \State \quad Assert that $P_{\text{best\_fixed\_current}} = \Pnewonchainfixed\_{\text{from\_event\_if\_accepted}}$.
    \Else
        \State \textbf{Verification 2.2 (Rejected Update - State Unchanged):}
        \State \quad Retrieve current parameters from $\CPoIm$: $\Thetabestfixedcurrent$.
        \State \quad Retrieve current metrics from $\CPoIm$: $P_{\text{best\_fixed\_current}}$.
        \State \quad Assert that $\Thetabestfixedcurrent = \Thetabestfixedold$.
        \State \quad Assert that $P_{\text{best\_fixed\_current}} = P_{\text{best\_fixed\_old}}$.
    \EndIf
    \State \Return accepted.
\end{algorithmic}
\end{algorithm}

\vspace{-5pt}

\begin{algorithm}
\caption{Part 3: TransferAndVerifyL1InferenceModel$(\CPoIm, \CInfer)$}
\label{alg:part3_transfer_l1_final}
\begin{algorithmic}[1]
    \State Retrieve best parameters from $\CPoIm$:
    \State \quad $L_{\text{L2}} \leftarrow \CPoIm.\text{getCurrentModelParameters}().\text{layerSizes}$.
    \State \quad $W_{\text{L2}} \leftarrow \CPoIm.\text{getCurrentModelParameters}().\text{weights}$.
    \State \quad $b_{\text{L2}} \leftarrow \CPoIm.\text{getCurrentModelParameters}().\text{biases}$.
    \State \quad $S_{w,\text{L2}} \leftarrow \CPoIm.\text{getScalingFactorWeights}()$.
    \State Let $\Theta_{\text{from\_L2}} \leftarrow \{W_{\text{L2}}, b_{\text{L2}}, L_{\text{L2}}\}$.
    \State Call $\CInfer.\text{setModelParameters}(\Theta_{\text{from\_L2}}, S_{w,\text{L2}})$ on the L1 network to update $\CInfer$.
    \State \textbf{Verification 3 (L1 Parameter Consistency after Transfer):}
    \State \quad Retrieve parameters from $\CInfer$:
    \State \quad \quad $L_{\text{L1}} \leftarrow \CInfer.\text{getLayerSizes}()$, $W_{\text{L1}} \leftarrow \CInfer.\text{getWeights}()$, $b_{\text{L1}} \leftarrow \CInfer.\text{getBiases}()$.
    \State \quad \quad $S_{w,\text{L1}} \leftarrow \CInfer.\text{getScalingFactorWeights}()$.
    \State \quad Assert that $\{W_{\text{L1}}, b_{\text{L1}}, L_{\text{L1}}\}$ equals $\Theta_{\text{from\_L2}}$.
    \State \quad Assert that $S_{w,\text{L1}}$ equals $S_{w,\text{L2}}$.
\end{algorithmic}
\end{algorithm}


\begin{algorithm}[H]
\caption{Part 4: VerifyL1OnChainInference$(\Xsample, \Moffchain^{\text{L1}}, \CInfer)$}
\label{alg:part4_verify_l1_inference_final}
\begin{algorithmic}[1]
    \State Retrieve $W_{\text{L1}}, b_{\text{L1}}, L_{\text{L1}}, S_{w,\text{L1}}$ from $\CInfer$ (as per getters in Part 3).
    \State Configure $\Moffchain^{\text{L1}}$ with parameters $\fromfixed{W_{\text{L1}}}{S_{w,\text{L1}}}$, $\fromfixed{b_{\text{L1}}}{S_{w,\text{L1}}}$, and $L_{\text{L1}}$.
    \State $\Yoffchain \leftarrow \Moffchain^{\text{L1}}.\text{predict}(\Xsample)$.
    \State $\Xsamplefixed \leftarrow \tofixed{\Xsample}{S_{w,\text{L1}}}$.
    \State $\Yonchain \leftarrow \CInfer.\text{classify}(\Xsamplefixed)$.
    \State \textbf{Verification 4 (Inference Output Consistency):}
    \State Assert that $\Yonchain$ equals $\integercast{\Yoffchain}$.
\end{algorithmic}
\end{algorithm}

\end{document}